\documentclass[%
 aip,
 pop,
 amsmath,amssymb,floatfix,
 reprint,%
]{revtex4-1}

\usepackage{graphicx}
\usepackage{dcolumn}
\usepackage{bm}

\usepackage[utf8]{inputenc}
\usepackage[T1]{fontenc}
\usepackage{mathptmx}
\usepackage{etoolbox}

\makeatletter
\def\@email#1#2{%
 \endgroup
 \patchcmd{\titleblock@produce}
  {\frontmatter@RRAPformat}
  {\frontmatter@RRAPformat{\produce@RRAP{*#1\href{mailto:#2}{#2}}}\frontmatter@RRAPformat}
  {}{}
}%
\makeatother

\begin{document}


\title{Burn Propagation in Magnetized High-Yield Inertial Fusion} 



\author{S. T. O'Neill}
\email{sam.oneill15@imperial.ac.uk}
\altaffiliation[Currently at ]{York Plasma Institute, University of York, Heslington YO10 5DD, United Kingdom}
\affiliation{The Centre for Inertial Fusion Studies, The Blackett Laboratory, Imperial College, London SW7 2AZ, United Kingdom}

\author{B. D. Appelbe}
\affiliation{The Centre for Inertial Fusion Studies, The Blackett Laboratory, Imperial College, London SW7 2AZ, United Kingdom}
\author{A. J. Crilly}
\affiliation{The Centre for Inertial Fusion Studies, The Blackett Laboratory, Imperial College, London SW7 2AZ, United Kingdom}
\author{C. A. Walsh}
\affiliation{Lawrence Livermore National Laboratory, P.O. Box 808, Livermore, California 94551-0808, USA}
\author{D. J. Strozzi}
\affiliation{Lawrence Livermore National Laboratory, P.O. Box 808, Livermore, California 94551-0808, USA}
\author{J. D. Moody}
\affiliation{Lawrence Livermore National Laboratory, P.O. Box 808, Livermore, California 94551-0808, USA}
\author{J. P. Chittenden}
\affiliation{The Centre for Inertial Fusion Studies, The Blackett Laboratory, Imperial College, London SW7 2AZ, United Kingdom}


\date{\today}

\begin{abstract}
Recent experiments at the National Ignition Facility (NIF) have demonstrated ignition for the first time in an inertial confinement fusion (ICF) experiment \cite{theindirectdriveicfcollaborationAchievementTargetGain2024}, a major milestone allowing the possibility of high energy gain through burn propagation. Use of external magnetic fields, applied primarily to reduce thermal losses, could increase hotspot temperature and ease requirements for ignition, opening up the capsule design space for high energy gain. However, this same restriction of thermal transport has the potential to inhibit burn propagation \cite{jonesPhysicsBurnMagnetized1986}, which is vital in the attainment of high gain. In this work, radiation-magnetohydrodynamics (MHD) simulations carried out using the code Chimera are used to investigate the effect of a pre-imposed magnetic field on ignition and burn propagation. This paper studies the propagation of burn using both an idealized planar model and in fully-integrated 2D MHD simulations of an igniting NIF capsule. A study of magnetised burn propagation in the idealized planar model identifies three regimes of magnetized burn propagation: (1) thermal conduction driven; (2) alpha transport driven; and (3) fully suppressed burn. Simulations of NIF shot N210808 with an applied 40T axial field show clear indication of burn suppression perpendicular to field lines, with rapid burn observed along field lines. Implosion shape is altered by the field, and anisotropic conduction causes significant modification to the rate of ablation during stagnation. These results highlight the fundamental changes to implosion dynamics in high-yield magnetized ICF and motivate further study to better optimize future magnetized target designs for high gain.
\end{abstract}


\maketitle 


\section{Introduction}
\label{sec:intro}

Recent experiments at the National Ignition Facility (NIF) \cite{mosesNationalIgnitionFacility2005} have demonstrated for the first time that target energy gains greater than 1 are possible on current laboratory indirect drive inertial confinement fusion (ICF) experiments \cite{indirectdriveicfcollaborationLawsonCriterionIgnition2022,theindirectdriveicfcollaborationAchievementTargetGain2024}. This achievement is made possible by the obtainment of ignition in so-called `hotspot ignition' designs, where the energy deposited by fusion alpha particles within the central hotter and lower density region (the `hotspot') of the compressed deuterium-tritium (DT) fuel assembly dominates energy losses, causing a rapid rise in temperature. Energy losses from the hotspot consist of thermal conduction loss into the cold, dense DT fuel shell surrounding the capsule; escape of $\alpha$ particles from the hotspot; radiative losses; and mechanical work exerted by the hotspot on the confining shell mass. Simple energy balance models for the ignition condition \cite{lindlDevelopmentIndirectDrive1995} suggest that to obtain ignition hotspot temperatures of $> 5$~keV and densities $\rho > 30$~g~cm$^{-3}$ are required. A cold fuel shell with $\rho> 300$~g~cm$^{-3}$ and $T < 500$~eV is required for adequate confinement \cite{atzeniPhysicsInertialFusion2008}, which is typically formed by $\sim 30$-fold convergence of a spherical target to form a hotspot of radius $R_{HS} \sim 50$~$\mu$m.

The onset of ignition in any hotspot ignition ICF experiment leads to the propagation of a thermonuclear burn wave into the dense fuel shell \cite{lindlDevelopmentIndirectDrive1995,christophersonThermonuclearIgnitionOnset2019}, and signatures of burn propagation have recently been observed experimentally \cite{crillyMeasurementsDenseFuel2024b,ruberyHohlraumReheatingBurning2024a}. The efficiency of this process is vital for the obtainment of high nuclear yields, as $> 90\%$ of the total DT fuel mass is contained in this layer. However, the rapidly rising central pressure due to hotspot ignition drives rapid disassembly of the fuel capsule, which quenches burn propagation and leads to short timescales available for burn. The burn wave is driven by the same energy transport processes from the hotspot into the cold fuel outlined above, which although detrimental for ignition, are highly significant for burn propagation \cite{tongBurnRegimesHydrodynamic2019}. Most important is the varying role of alpha particle deposition both spatially and temporally during burn. Deposition by DT $\alpha$ particles, born with energy $E_\alpha \sim 3.54$~MeV, mainly occurs due to coulomb scattering with electrons, where the birth alpha particle mean free path for $\alpha-e$ collisions can be calculated for a 50:50 DT plasma using \cite{helanderCollisionalTransportMagnetized2002},
\begin{equation}\label{eq:alpha_range}
    l_{\alpha e} = 2.61 \times 10^{-12}  \frac{ T_e^{3/2}}{\rho_{DT} \ln \Lambda_{\alpha e}} \text{ m}\,,
\end{equation}
where $T_e$ is the electron temperature in eV, $\rho_{DT}$ is the fuel mass density in kg~m$^{-3}$, and $\ln \Lambda_{\alpha e}$ is the Coulomb logarithm for $\alpha -e$ collisions. This indicates the strong variation in alpha particle range moving radially from the hotspot to the cold, dense fuel shell, with typical stopping lengths of $l_{\alpha e}^{HS}\sim 50$~$\mu$m~$\sim R_{HS}$ in the central hotspot, falling to $l_{\alpha e}^{sh}\sim 0.1$~$\mu$m in the shell. This means that alpha particles will act to transport energy non-locally from their birth in the central hotspot, before rapidly depositing their energy as they reach the cold fuel interface (i.e. the burn front).

A complementary but alternative approach to the experiments above is the addition of an external magnetic field to the ICF target before compression, the presence of which can restrict energy transport by charged particles moving perpendicular to field lines. Experiments utilizing external magnetic fields, often termed \textit{magnetized ICF}, include a broad range of inertial fusion schemes, such as: pulsed-power driven targets (most notably the MagLIF experiments at Sandia National Laboratories \cite{slutzHighGainMagnetizedInertial2012,yager-elorriagaOverviewMagnetoinertialFusion2022}); direct drive laser fusion \cite{knauerCompressingMagneticFields2010,changFusionYieldEnhancement2011,boseEffectStronglyMagnetized2022}; and indirect drive laser fusion \cite{moodyTransientMagneticField2020,moodyIncreasedIonTemperature2022,sioPerformanceScalingApplied2023} which will be the main focus of this study. The primary benefit of the magnetic field in regards to indirect drive hotspot ignition schemes is the suppression of electron thermal conduction losses into the shell, which lowers the $\rho R_{HS}$ requirement for ignition, along with increasing the bulk hotspot temperature compared to an equivalent implosion without field \cite{walshMagnetizedICFImplosions2022}. Current magnetized ICF experiments at the NIF make use of a magnetic field applied axially along the hohlraum generated by a pulsed-power coil \cite{moodyTransientMagneticField2020}. Experiments to date\cite{moodyIncreasedIonTemperature2022,sioPerformanceScalingApplied2023,strozziDesignModelingIndirectlyDriven2024} have made use of initial magnetic fields with initial strengths of up to 28~T on room-temperature deuterium ($D_2$) targets, and have demonstrated ion temperature enhancement of $\sim 1$~keV and fusion yield enhancement of $\times 3$ in this relatively low-yield platform.

During the implosion phase the dominant magnetic field transport mechanism can be estimated using the magnetic Reynolds number, the ratio of magnetic advection to magnetic diffusion, $R_M \sim UL/\eta$ where $U$ is the fluid velocity, $L$ is the implosion scale length and $\eta$ is the magnetic diffusivity. Due to the high characteristic fluid velocities associated with ICF implosions $R_M$ is large, meaning that magnetic flux is frozen-in to the plasma during compression. For typical ICF convergence ratios this leads to expected peak stagnation field strengths of $> 10$~kT for an initial axial field $\sim 30$~T. The impact of magnetization on electron thermal and magnetic transport is quantified by the \textit{Hall parameter}, $\omega \tau$, the product of the electron gyrofrequency ($\omega_e$) and the electron collision time ($\tau_e$), given for a DT plasma by,
\begin{equation}\label{eq:hall}
    \omega \tau = 2.51 \times 10^{-4} \frac{T_e^{3/2} \left|\mathbf{B}\right|}{\rho_{DT} \ln \Lambda_{ei}} \,,
\end{equation}
where $\mathbf{B}$ is the magnetic field in T and $\ln \Lambda_{ei}$ is the Coulomb logarithm for electron-ion collisions. From Braginskii \cite{braginskiiTransportProcessesPlasmas1965}, electron thermal conductivity becomes significantly altered when $\omega \tau \sim 1$, with the thermal conductivity in the direction perpendicular to the magnetic field ($\kappa_\perp$) reduced by $\sim 67\%$ the value compared to along the direction of the field ($\kappa_\parallel$). Using the typical conditions in an ICF hotspot along with the estimate for compressed field strengths, peak Hall parameters of $\sim 10$ could be possible in the hotspot, which corresponds to $\kappa_\perp/\kappa_\parallel \sim 0.01$. Similarly to $l_{\alpha e}$, the Hall parameter varies rapidly with density and temperature, meaning that generally $\omega \tau$ will decrease rapidly moving radially towards the fuel shell.

In addition to suppression of electron thermal conductivity, the addition of the magnetic field restricts the transport of alpha particles perpendicular to magnetic field lines to the alpha Larmor radius, $r_{L \alpha}$. For a 3.54~MeV DT fusion $\alpha$ particle, $r_{L}^\alpha \sim R_{HS} \sim l_{\alpha e}$ when $B \sim 5$~kT, meaning that the fraction of alpha particle energy deposited in the hotspot will be expected to increase for the field strengths predicted in a magnetized ICF hotspot. The combined effect of thermal conduction suppression and alpha particle confinement can inhibit energy losses from the hotspot, lowering ignition requirements.

There are, however, complications to the use of magnetic fields in spherical ICF implosion designs, of which two will be considered in this paper. Firstly, the use of an axial field in a ICF implosion adds an inherent perturbation to the spherical implosion geometry, with different energy transport rates perpendicular and parallel to magnetic field lines, i.e. along the capsule equator versus the capsule pole. One such impact of this anisotropy, on implosion drive and bulk hotspot shape in magnetized room-temperature capsule simulations, has been observed in previous studies \cite{walshPerturbationModificationsPremagnetisation2019,walshMagnetizedICFImplosions2022}, where it is seen that capsule shape is elongated along the direction of applied field, corresponding to a measured mode-2 Legendre polynomial shape ($P_2$) as inferred from synthetic x-ray images. Analysis of magnetized ICF experiments \cite{strozziDesignModelingIndirectlyDriven2024} show variation in implosion shape with applied field, however interpretation of this is complicated by shot-to-shot variations in laser drive and hohlraum drive asymmetry, which are not considered in the current work. In general, low-mode shape perturbations to the implosion have a significant impact on fusion performance if not correctly mitigated \cite{caseyDiagnosingOriginImpact2023}. Secondly, and of primary importance in this paper, is the possible suppression of burn propagation in the presence of a magnetic field due to magnetic insulation of the hotspot. Previous works \cite{jonesPhysicsBurnMagnetized1986,velikovichThermonuclearBurnWave2012,appelbeMagneticFieldTransport2021} have studied the reduction in burn rate due to magnetic fields, however these studies did not comprehensively model many of the aspects of the physics relevant to axially magnetized spherical ICF targets, whilst previous integrated magnetized capsule simulations \cite{perkinsTwodimensionalSimulationsThermonuclear2013,perkinsPotentialImposedMagnetic2017} did not study burn propagation in detail.

In this work these two issues are studied in detail for the first time during the stagnation, ignition and burn phases using magnetohydrodynamics simulations representative of current high-yield NIF implosion designs. All simulations in this paper make use of Chimera, a 3D Eulerian radiation-magnetohydrodynamics code with in-line Monte Carlo alpha particle stopping and burn model \cite{chittendenRecentAdvancesMagnetohydrodynamic2009,Chittenden2016a,mcglincheyDiagnosticSignaturesPerformance2018,walshPerturbationModificationsPremagnetisation2019,tongBurnRegimesHydrodynamic2019}. Chimera has extended-MHD capabilities with anisotropic heat flow \cite{walshExtendedmagnetohydrodynamicsUnderdensePlasmas2020} using Epperlein and Haines transport coefficients \cite{epperleinPlasmaTransportCoefficients1986}. Radiation transport utilizes a $P_{1/3}$ automatic flux-limited scheme \cite{olsonDiffusionP1Other2000,mihalasTimedependentRadiativeTransfer1982} with tabulated atomic data from the code SpK \cite{crillySpKFastAtomic2023}. Equation of state is handled by reading in tabulated data from Frankfurt Equation of State (FEOS) \cite{faikEquationStatePackage2018}. Primary DT fusion alphas are modeled using an inline magnetized alpha particle transport and burn package \cite{tongBurnRegimesHydrodynamic2019}, with a Monte Carlo stopping model \cite{sherlockMonteCarloMethodCoulomb2008} using stopping powers from either Spitzer \cite{spitzerPhysicsFullyIonized2006} or Maynard-Deutsch-Zimmerman (MDZ) \cite{maynardBornRandomPhase1985,zimmermanMonteCarloMethods1997}.

Section~\ref{sec:planar_burn} studies magnetized burn physics in an idealized model in 1D planar geometry with a uniform initial field profile. This model allows for a clearer study of burn dynamics without conflicting impacts of the field on other aspects of capsule implosion performance, and allows for verification of previous work on this topic \cite{appelbeMagneticFieldTransport2021} using the full multi-physics capabilities of Chimera. Using this model, three distinct regimes of magnetized burn propagation are identified. The first, at low Hall parameters, involves non-local alpha heating being aided by electron thermal conduction, which transport heat into the dense fuel shell. The second regime corresponds to large Hall parameters ($\omega \tau \sim 10$), where energy transport from the hot fuel is dominated by non-local alpha heating, and shows qualitatively different burn dynamics and a large decrease in the rate of burn. Finally, at very large electron Hall parameters ($\sim 100$), alpha particles are also magnetically confined, and burn propagation is almost completely suppressed.

The results in Section~\ref{sec:planar_burn} map to the two extreme regimes of burn propagation expected to be found in a magnetised capsule implosion: the propagation of burn along field lines in a capsule, along the capsule pole, equivalent to the unmagnetized case; and the propagation of burn perpendicular to field lines, corresponding to burn at the capsule equator. Section~\ref{sec:capburn} presents results from fully-integrated 2D MHD simulations based on the design for NIF shot N210808 \cite{kritcherDesignInertialFusion2022}, which ignited with a fusion yield of 1.3~MJ. These simulations are then carried out with a 40~T initial applied axial magnetic field throughout the implosion. Impact of the field on the shape of the cold fuel shell during stagnation are studied in Section~\ref{subsec:hotspot}, before detailed study of burn propagation along field lines in Sec.~\ref{subsec:burn} for the two extreme cases, along the capsule pole and equator, are studied and compared to the results found in the 1D planar model. In both of these cases the regime of burn obtained perpendicular to magnetic field lines is markedly different from that along field lines (or without field), and indicates that burn propagation is suppressed at the capsule equator.

\section{Modeling Magnetized Burn in a 1D Planar Geometry}
\label{sec:planar_burn}

\subsection{Model Details}\label{sec:3_model}
\begin{figure}
    \centering
    \includegraphics{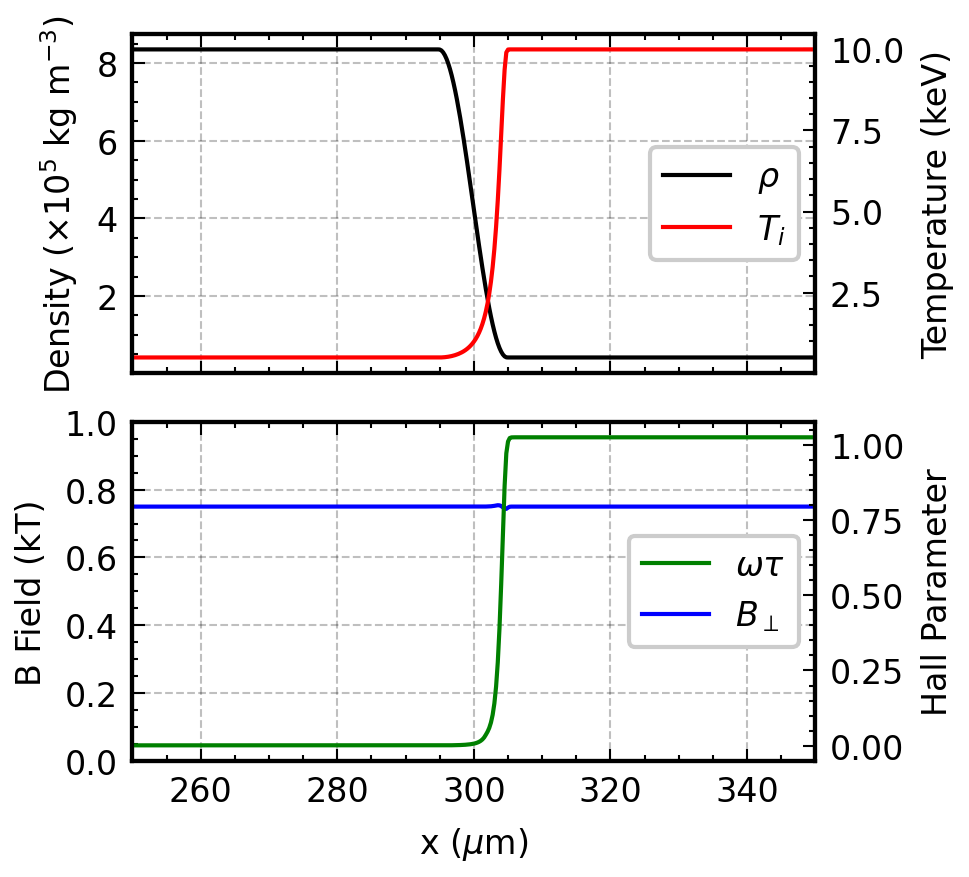}
    \caption{Typical initial conditions in 1D for the planar burn model. Top: Density (Black) and temperature (red) profiles with hot fuel on the left and cold fuel on the right, separated by a 10~$\mu$m sigmoidal transition region. Bottom: Initially uniform perpendicular magnetic field is imposed (blue), giving a Hall parameter profile (green) which peaks in the hot fuel.}\label{fig:planar_initial}
\end{figure}

Simulations in this section are initialized with isobaric, semi-infinite reservoirs of hot, low density and cold, high density 50:50 DT plasma with a finite sigmoidal transition region between them as shown in the top panel of Figure~\ref{fig:planar_initial}. Here, the gradients in temperature and density are in the x-direction. Throughout this section, the hot fuel region will be referred to as the `hotspot', to aid in comparison to capsule simulations in Section~\ref{sec:capburn}. Similarly, the cold dense fuel region will be referred to as the `cold fuel'.
 
An initially uniform magnetic field is imposed in the perpendicular direction ($+z$ direction), which, although not exactly representative of the field profile in a stagnating ICF capsule aids in interpretation of results, whilst the consequences of a realistic initial field profile are studied in Section~\ref{sec:capburn}. Due to the strong temperature and density dependence of the electron Hall parameter (Eq.~\ref{eq:hall}) this produces a large $\omega \tau$ in the hot, low density fuel region which decreases rapidly moving into the cold fuel. The initial Hall parameter is quoted as $\omega \tau_0$, the initial magnetization in the hot fuel. 
 
Initial density and temperature is based on a representative burning NIF capsule, with initial density and temperature in the hot fuel $n_0 = 10^{31}$~m$^{-3}$, $T_0 = 10$~keV and initial cold fuel temperature of $T_0^c = 500$~eV, with a finite transition between the two regions with length $l_T = 10$~$\mu$m. The density of the cold fuel region is set such that initial conditions are isobaric. The simulation domain is 2400 cells with a resolution 0.25~$\mu$m, sufficient such that transient waves do not reflect off boundaries and interfere with the physics of interest, and such that results have converged  (i.e. length scales required for burn propagation are resolved). Other initial conditions have been studied, such as lower initial densities and larger transition scale lengths. Although not presented here, these have shown qualitatively similar behaviour, showing that the underlying burn physics should be relevant across a broad range of ICF hotspot parameters and indeed other potential magneto-inertial fusion schemes.

Simulations in planar geometry are carried out using Chimera, as described in Section~\ref{sec:intro}. All results use a fully-ionised, ideal gas EoS model. Flux-limited thermal conduction is used with an electron flux-limiter of 0.1 and ion flux limiter of 0.5.  Radiation transport is carried out using 10 frequency groups, binned between 0.1~eV and 30~keV, which is adequate to resolve the radiation transport of interest here. The results obtained include all extended-MHD terms relevant in 1D. The alpha model uses MDZ stopping powers, where the number of alpha macro-particles has been chosen to ensure convergence within the simulations.

The hydrodynamic boundary conditions are set so that the left-hand (cold fuel) boundary is transmissive/outgoing and the right-hand (hotspot) boundary is reflective. The flux of alpha particles and radiation are suppressed by a constant factor of 6 to account for the lack of spherical divergence in planar geometry, accounting for the infinite extent, and a corresponding overestimation of non-local heating effects. This can be justified by writing the flux ($\Phi$) as,
\begin{flalign}
    && &\Phi \propto \left(\frac{\text{Volume}}{\text{Surface Area}}\right) \times \text{Source Term} &&\\
    && &\left(\frac{V}{S}\right)_{sph} = \frac{\frac{4}{3}\pi r^3}{4 \pi r^2} = \frac{r}{3} &&\\
    && &\left(\frac{V}{S}\right)_{planar} = \frac{r^2L}{r^2} = 2r \label{eq:v_s_planar} &&
\end{flalign}
where we have used $L = 2r$ in Equation~\ref{eq:v_s_planar} due to the reflective boundary condition. This implies that the effective spherical flux is $\Phi_{sph} = \Phi_{planar}/6$. The inclusion of this dilution factor brings the magnitude of non-local alpha and radiation transport back into the regime expected from capsule simulations. Use of this dilution factor allows results obtained to be compared to those in Section~\ref{sec:capburn}.

One numerical feature in these simulations is the formation of a pressure release wave on initialisation, due to the large temperature gradient present in the initial conditions. Thermal conduction occurs across this interface more rapidly than the fluid can respond hydrodynamically, which perturbs the initial pressure profile, driving this release wave. This wave travels rapidly in the hot, low density fuel, meaning it is not visible in the results here. The wave emitted in the direction of the cold fuel is noticeable and is visible as a density spike moving in the negative $x$-direction. Although this feature is visible in the results here, it does not have a significant impact on the conclusions presented as it quickly moves away from the region of interest. 

The geometry and initial conditions here are selected to allow the study of magnetized burn independent of the complications associated with realistic magnetized target designs, allowing for a clearer picture of transport processes at magnetized burn fronts to be built. The dynamics of burn in a realistic capsule geometry will be studied in Section~\ref{sec:capburn}.

\subsection{Unmagnetized Burn}

\begin{figure}
    \centering
    \includegraphics{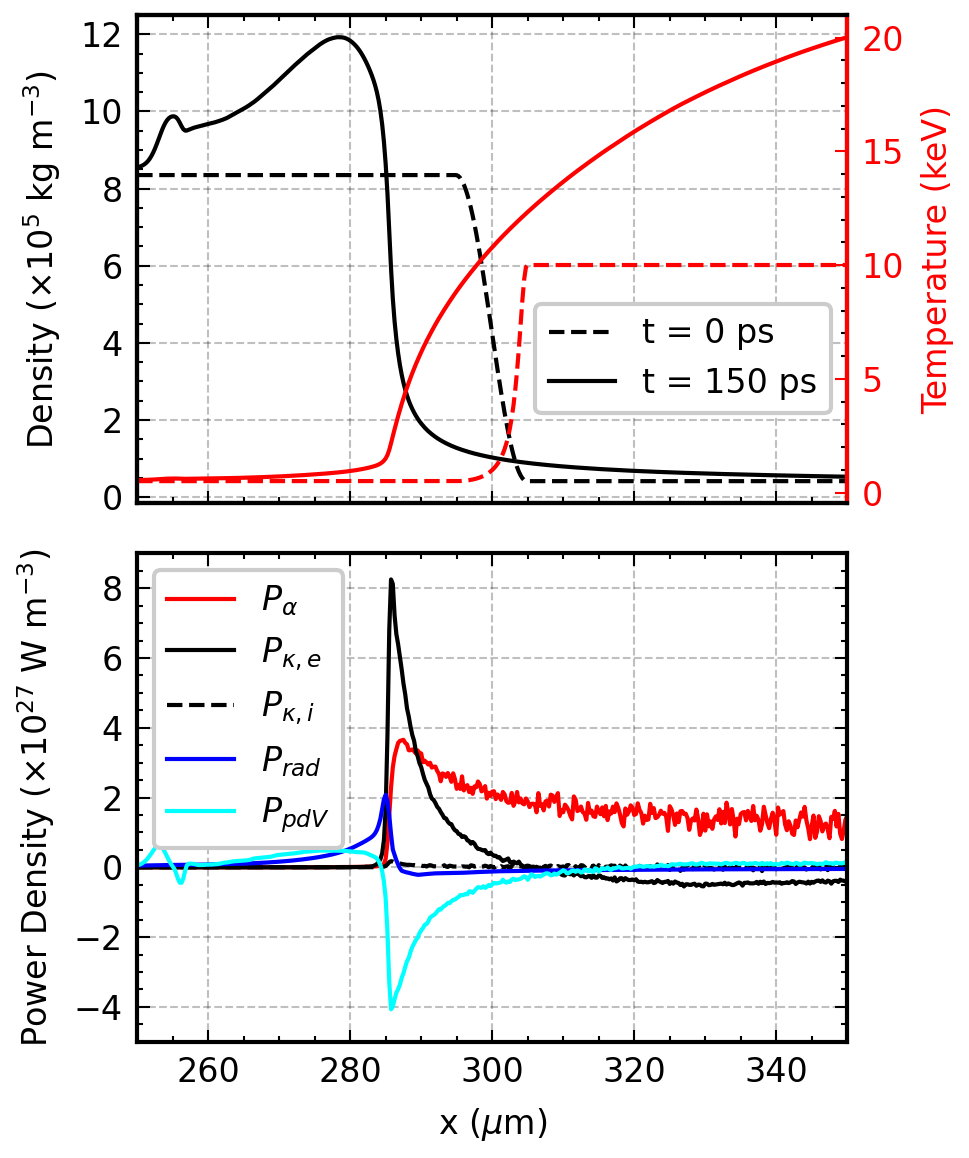}
    \caption{Top: Density (black) and ion temperature (red) profiles after 150~ps of burn propagation with no field applied, alongside the initial profiles (dashed). Bottom: Contributions to the power balance equation 150 ps into burn, shown as a power density. $P_\alpha$ is the alpha heating (red); $P_{\kappa,e}$ is the electron thermal conduction (black); $P_{\kappa,i}$ is the ion thermal conduction (black dashed); $P_{rad}$ is the radiation transport (blue); and $P_{PdV}$ is the hydrodynamic work (cyan)}
    \label{fig:planar_unmag-burn}
\end{figure}

Initially, it is useful to study how burn propagates in the planar model with no magnetic field present. The top panel of Figure~\ref{fig:planar_unmag-burn} shows the density and temperature profiles 150~ps into burn propagation. The bottom panel shows the relevant terms in the power balance equation at the final timestep. In the hotspot, the central temperature rises rapidly from 10~keV initially to $\sim 20$~keV over the 150~ps of burn considered, due to the strong alpha heating across this region. During this time the temperature profile smooths significantly due to electron thermal conduction, observed as a negative power contribution from thermal conduction across the hotspot region. This results in a rapid relaxing of the temperature gradient length scale, and within 150~ps the initial sharp $10$~$\mu$m transition has relaxed into a smooth temperature gradient across $> 50$~$\mu$m. 

At the interface between hot and cold fuel, propagation of the burn front is observed. Here the 2~keV ion temperature contour is used as a measure for the position of the burn front, which propagates $\sim 15$~$\mu$m into the cold fuel over 150~ps. As energy is transported into the cold fuel the pressure rapidly increases. On the high temperature side of the burn front, this causes cold material to stream in, and this material is heated. This introduction of colder fuel reduces the temperature in the hotspot closest to the burn front, observed as the negative spike in $PdV$ work. In this region the fuel density rises, forming a region of relatively dense ($> 100$~kg~m$^{-3}$) and hot ($> 2$~keV) fuel. Over time this fuel is incorporated into the hot fuel region and, as fusion reactivity scales with $n^2$, gives a large boost to fusion yields and therefore hotspot temperatures through what is termed `bootstrap heating'.

On the low temperature side of the ablation front, material expands into the cold fuel region. As this fuel is expanding towards a large reservoir of cold, dense fuel it is quickly slowed down, causing a pile-up of mass in the cold fuel region. In the context of an ICF capsule, this corresponds to an infinitely confined hotspot. In a realistic capsule geometry, this process instead corresponds to re-expansion of the fuel assembly, highlighting an important limitation of this model compared to full spherical ICF simulations. Despite this, the burn exhibited here does operate in a relevant regime to capsule experiments. For an igniting ICF target design, such as N210808, simulations show \cite{kritcherDesignInertialFusion2022} that a typical burn width is $\sim 100$~ps, central hotspot temperature $\sim 10 - 20$ keV and propagation of burn across a few tens of microns of cold fuel (see Table~\ref{tab:cap_metrics0T}). These values are of similar magnitude to the spatial, temperature and time scales measured in this planar set-up. 

The power density profiles in the bottom panel of Figure~\ref{fig:planar_unmag-burn} illustrate the dominant processes driving the propagation of burn in this system, particularly the complementary roles of thermal conduction and alpha heating at the burn front. Alpha heating provides an approximately uniform heating source across the main region of the hotspot ($x > 300$~$\mu$m), which dominates over loss terms, leading to increasing fuel temperature with time. Within this hot, low density fuel region the alphas will have a long mean free path. Using Equation~\ref{eq:alpha_range}, the estimated birth alpha mean free path is $l_\alpha \sim 180$~$\mu$m using the initial hot fuel parameters. As this is of similar magnitude to the length of the hotspot considered here, this means that alpha particles will stream relatively freely across this region, only depositing a fraction of their energy. When the alpha particles reach the cold, dense fuel the mean free path drops rapidly. For the estimated temperature (2~keV) and density (100~kg~m$^{-3}$) in the ablation region $l_\alpha \sim 10$~$\mu$m, falling to $l_\alpha \sim 0.2$~$\mu$m in the conditions corresponding to the initially cold fuel. This means that as the alphas reach this region they rapidly slow and deposit their energy, corresponding to the peak in alpha heating seen at approximately 285~$\mu$m in Figure~\ref{fig:planar_unmag-burn}. This peak will be termed the `non-local alpha heating peak' in the remainder of this paper, and corresponds to the alpha particle Bragg peak discussed previously in literature \cite{tongBurnRegimesHydrodynamic2019,frenjeMeasurementsIonStopping2015,cayzacExperimentalDiscriminationIon2017}. 

Close to the burn front ($280 < x < 290$~$\mu$m), the largest contribution to the power balance is electron thermal conduction due to steep temperature gradients, with a peak deposited power density roughly twice that of alpha heating, albeit across a narrower region. The peak of the deposition due to thermal conduction is ahead of the non-local alpha heating peak, showing that thermal conduction rapidly transports energy deposited by the alpha heating further into the cold fuel. From this, a feedback mechanism for the propagation of burn into the cold fuel can be proposed. As alpha particles slow, they deposit their energy in a non-local heating peak close to the burn front. Electron thermal conduction carries this energy further into the cold fuel, heating it up rapidly and increasing its transparency to alpha particles (by Equation~\ref{eq:alpha_range}). This means that alphas can now propagate further into the cold fuel before thermalizing, moving the non-local alpha heating peak further forward. This feedback between electron thermal conduction and non-local alpha deposition is observed to be an important driver for burn.

Ion thermal conduction is also plotted here, but is entirely negligible in the unmagnetized case, as it is lower than electron thermal conduction by a factor $(m_i/m_e)^{1/2}$. Energy transport by radiation is plotted but it is not a dominant effect. In the hot fuel region, radiative emissions are low enough not to be the dominant cause of cooling, however they aid to smooth out temperature gradients within the hotspot alongside electron thermal conduction. Close to the burn front, radiation deposition is observed, however, the peak magnitude is significantly lower than that due to either alpha heating or thermal conduction. The peak of this radiatively heated region is ahead of both the conduction and alpha heating peaks, and is significantly broader, meaning it mainly acts as a pre-heating mechanism for the cold fuel. The role of radiation in this regime of burn propagation is therefore to act as an additional source of heating ahead of the non-local alpha heating peak, of which a portion of this energy is recycled into the hot fuel by producing an enhancement to mass ablation rates. 
 
\subsection{Magnetized Burn}

Three magnetized simulations were carried out using the planar model, with initial field strengths of 750~T, 7500~T and 75,000~T respectively, intended to be representative of three distinct regimes of magnetized burn propagation. Characteristic values of the Hall parameter and alpha Larmor radius are given in Table~\ref{tab:planar_paras}. Here, the initial Hall parameter in the hot fuel increases by an order of magnitude in each case from $\sim 1$ to $\sim 100$. In all three cases the initial Hall parameter at the 2~keV contour is $<< 1$, indicating that thermal conductivity is not suppressed locally to the burn front in any of the cases. The peak alpha Larmor radius decreases as field strength increases, where in the case of largest initial field $r_L^\alpha < l_T$, implying that alpha particles will be magnetically confined within the hotspot. These three cases therefore correspond to regimes where: (1) thermal conduction is moderately suppressed in the hotspot (by $\sim 67\%$), but alpha particles are unconfined; (2) thermal conduction in the hot fuel is almost entirely suppressed, but alpha particles are only moderately confined; and (3) both thermal conduction and alpha transport is suppressed.

\begin{table}
    \centering
    \caption{Initial and final values of the peak Hall parameter ($\omega \tau^{HS}$), the Hall parameter at the 2~keV contour ($\omega\tau^{bf}$) and the minimum Larmor radius of an alpha particle at birth ($r_L^\alpha$) for the three field strengths considered.  }
    \label{tab:planar_paras}
    \begin{ruledtabular}
    \begin{tabular}{ccccc}
         $B_0$ (T) &$t_{burn}$ (ps)&$\omega\tau^{HS}$& $\omega\tau^{bf}$& $r_L^\alpha$ ($\mu$m)\\
        \hline
        750 &0& 1& 0.05& 358\\
        7500 &0& 10& $5 \times 10^{-3}$& 35.8\\
        75000 &0& 100& $5 \times 10^{-4}$& 3.58\\
        \hline
 750 & 150& 2.5& 0.02&345\\
 7500 & 150& 36& 0.2&28.5\\
 75000 & 150& 300& 2&2.75\\
    \end{tabular}
    \end{ruledtabular}
\end{table}

\begin{figure*}
    \centering
    \includegraphics{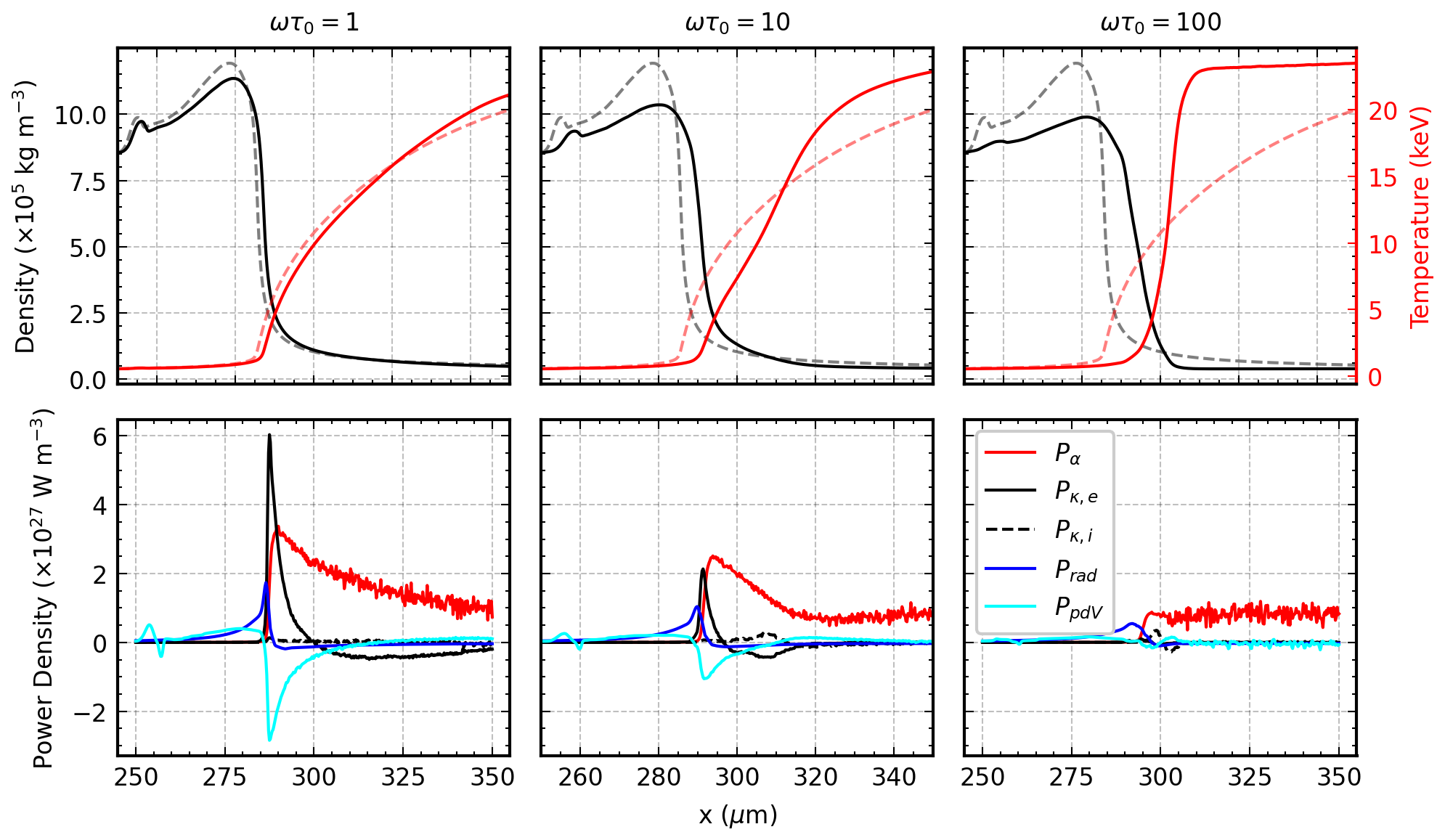}
    \caption{Profiles 150~ps into burn propagation for the three initial magnetizations considered showing: (Top) ion temperature in red and fuel density in black, compared against the case with no field present (dashed); (Bottom) Terms contributing to the power balance, plotted as a power density showing alpha deposition (red), electron thermal conduction (black), ion thermal conduction (black dashed), radiation (blue) and mechanical work (cyan).}
    \label{fig:planar_mag-burn}
\end{figure*}

Figure~\ref{fig:planar_mag-burn} shows the density and temperature profiles for three initial field strengths plotted alongside the unmagnetized case after 150~ps of burn propagation. Comparing the density and temperature profiles in the top panel, a distinct change in the nature of burn propagation is observed as the magnetic field strength increases. The $\omega \tau_0 = 1$ case is qualitatively identical to the unmagnetized case, with temperature profiles smoothed substantially by thermal conduction across the hot fuel. In the $\omega \tau_0 = 10$ case there is a more marked difference in the temperature and density profiles, with an inflection in the ion temperature profile between 300 - 310~$\mu$m producing a stepped ion temperature profile. This is caused by the almost complete suppression of heat flow at $x < 310$~$\mu$m, where  $\omega \tau > 2.5$. In this region, ion thermal conduction also plays a notable role due to the steepness of the temperature gradients, and the fact that the equivalent ion Hall parameter will be small. In the region $x > 310 $~$\mu$m, non-local alpha heating is significant, whilst $\omega \tau \sim 1$ or less, meaning that a burn wave can continue to develop.  In the case of $\omega \tau_0 = 100$, both energy transport processes out of the hotspot are suppressed meaning that burn propagation is substantially suppressed. In this case, the ion temperature profile statically self-heats with little change from the initial shape.

\begin{figure}
    \centering
    \includegraphics{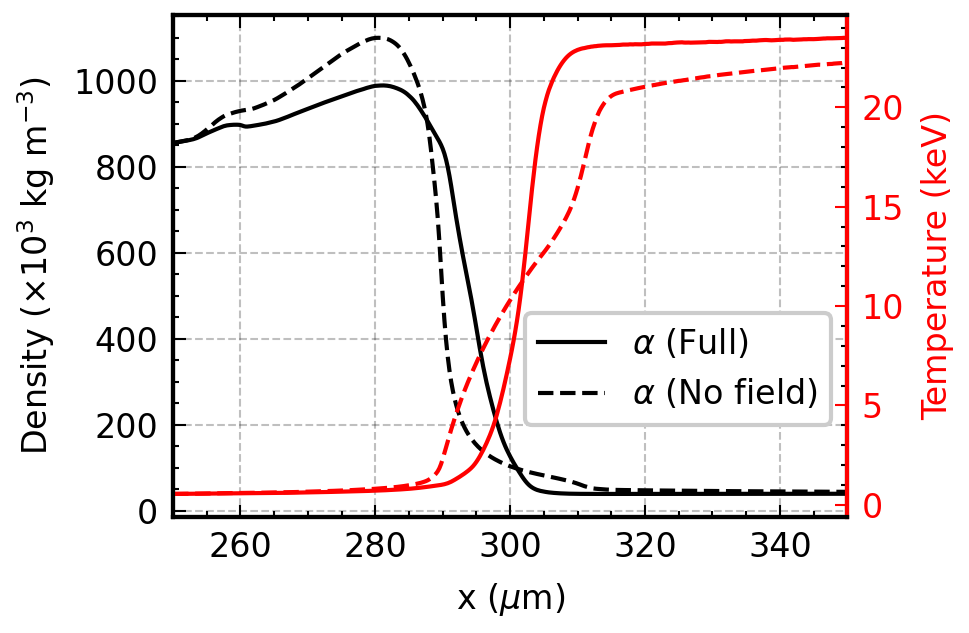}
    \caption{Density and temperature profiles for the $\omega\tau_0 = 100$ case after 150~ps comparing simulations with (solid) and without (dashed) magnetic field effects on the alpha particles.}
    \label{fig:planar_alpha-mag}
\end{figure}

An interesting comparison to consider is the case with $\omega \tau_0 = 100$ where magnetized $\alpha$ particle effects have been removed, as shown in Figure~\ref{fig:planar_alpha-mag}. Without alpha particle magnetization there is a return to the stepped temperature profile observed in the $\omega \tau_0 = 10$ case, marking a return to burn propagation. The magnitude of non-local alpha heating peak is higher than any of the previous cases, due to the much hotter and broader hot fuel region, which produces significantly more alpha particles. Higher temperatures also increase $l_{\alpha e}$, meaning a larger fraction of alpha energy reaches the burn front. This result indicates that non-local alpha heating is key in producing the regime of magnetized burn observed in the $\omega \tau_0 = 10$ case.

\begin{figure}
    \centering
    \includegraphics{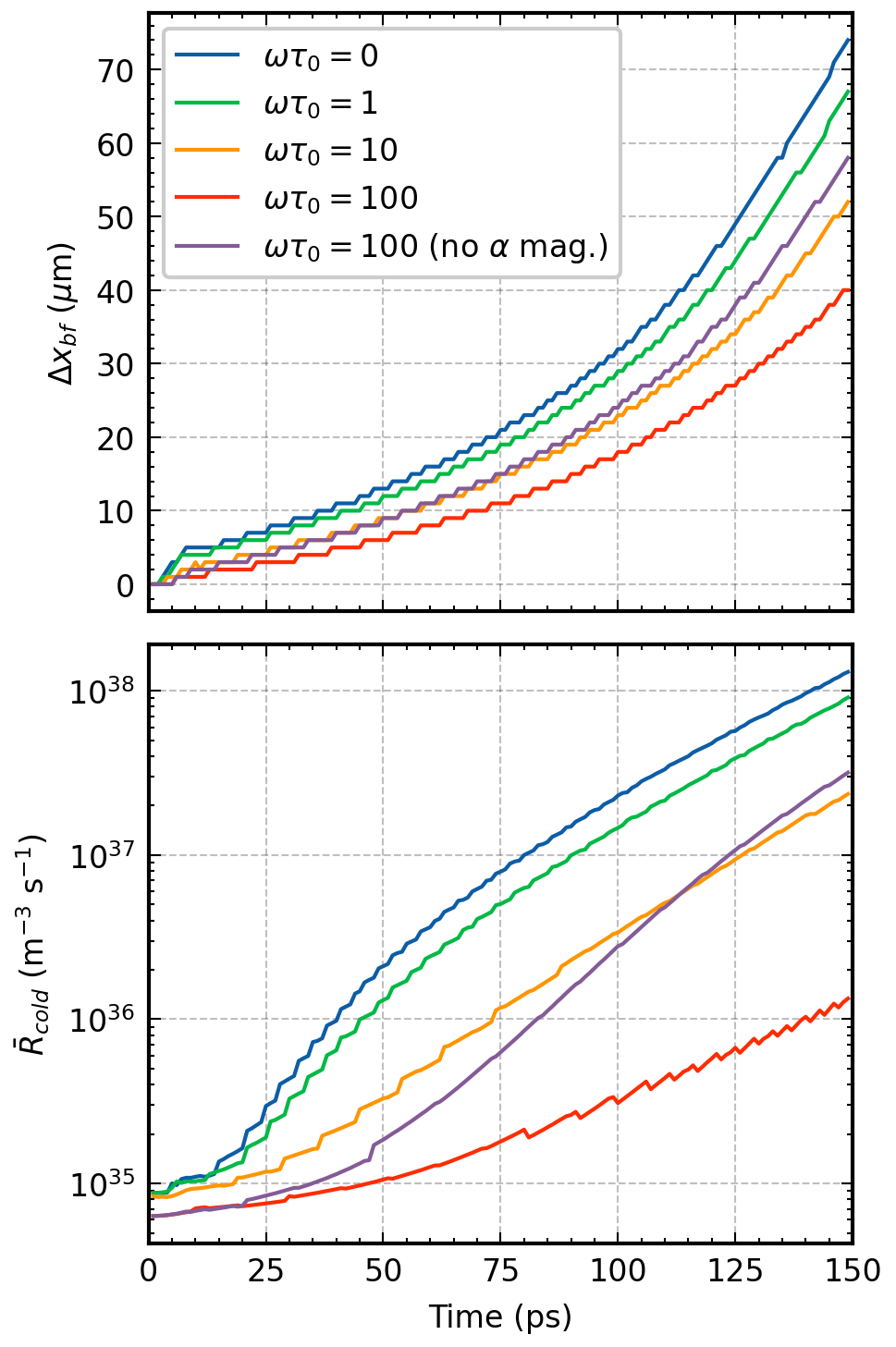}
    \caption{(Upper) Change in the position of the 2~keV contour (Lower) Volume averaged reactivity in all fuel initially $< 2$~keV as a function of time.}
    \label{fig:planar_diags}
\end{figure}

Quantitative measurement of the rate of burn propagation in this planar geometry is difficult, as typical ICF metrics such as total fusion neutron yield do not have meaning due to infinite extents. In Figure~\ref{fig:planar_diags}, two metrics for the impact of magnetization on the rate of burn propagation and consequence for possible neutron yield are presented. In the top panel, the position of the 2~keV ion temperature profile ($x_{bf}$) is plotted as a function of time, representative of the position of the burn front. The lower panel shows the volume averaged reactivity of the fuel that is initially at a temperature $< 2$~keV $\left( \bar{R}_{cold} = \frac{1}{x_{bf}} \int _0 ^{x_{bf}} \frac{1}{4} n^2 \left<\sigma v\right> dx\right)$. This quantity is important as it indicates the enhancement to total fusion rates purely through the propagation of burn, as is vital for the obtainment of high-gain ICF. Both metrics support the qualitative trends observed, with the rate of burn propagation and yield enhancement generally decreasing at higher initial field strengths. In particular, the $\omega \tau_0 = 0, 1$ cases show a rapid, exponential increase in cold fuel reaction rates early in time ($< 50$~ps), whereas other cases show a more gradual increase indicating a less efficient regime of burn propagation.  A notable exception to this is the highly magnetized case with alpha magnetization removed, which initially has burn rates similar to those in the full simulation, before transitioning into a regime of more rapid burn, overtaking the $\omega \tau_0 = 10$ case at $\sim 100$~ps. This shows that an efficient burn wave can be produced solely through the effect of non-local alpha heating, however this burn wave requires significantly more time to develop. This is due in part to the fact that heat transport solely by alpha particles is less efficient, and in part to the fact that alpha production rates are strongly dependent on hotspot temperature meaning that sufficient self-heating must occur before strong alpha driven burn is observed. In the case of the planar model, central temperatures $\gtrsim 20$~keV are required for this type of burn propagation. In the context of realistic ICF capsules, much longer confinement times would therefore be required in order to reach these high hotspot temperatures prior to capsule disassembly.

\subsection{Effect of magnetic field dynamics at the burn front}
\begin{figure}
    \centering
    \includegraphics[width=\linewidth]{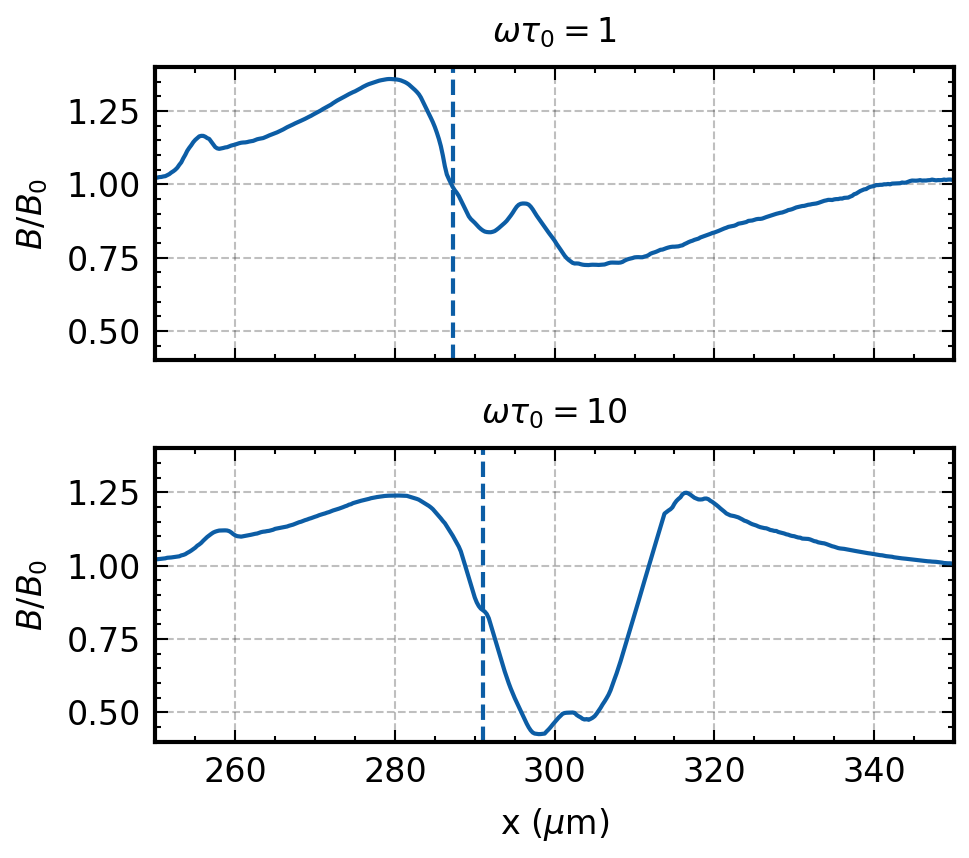}
    \caption{Magnetic field profiles normalised to the initial field strength after 150~ps of burn propagation for $\omega \tau_0 = 1,10.$ The position of the 2~keV ion temperature contour is marked by a dashed line in each case.}
    \label{fig:planar_field}
\end{figure}
In Table~\ref{tab:planar_paras} it is notable that the Hall parameter as measured locally at the burn front decreases over the course of burn for the $\omega \tau_0 = 1$ case, whilst increases for the cases with larger magnetisation due to magnetic field transport effects over the burn duration. The magnetic field profiles after 150~ps of burn for two runs are shown in Figure~\ref{fig:planar_field} normalized to the initial field magnitude. In both cases magnetic field advection frozen-in to the bulk fluid flow ($\mathbf{v_{fluid}}$) is dominant, and reduces the overall magnetic field at the burn front due to ablative flows. The major competing effect to this is Nernst advection, caused due to the effect of the magnetic field on the thermal force \cite{braginskiiTransportProcessesPlasmas1965}, which results in advection of the magnetic field down temperature gradients with velocity $\mathbf{v_N} = - \frac{\beta_\wedge}{e |\mathbf{B}|}\nabla T_e$, where the coefficient $\beta_\wedge$ is a function of $\omega \tau$ and is maximized at values of $\omega \tau \sim 1$.  This means that in the $\omega \tau_0 = 1$ case, the Nernst term is important, reducing the magnetic field strength in the hot fuel and compensating for the advection of magnetic field away from the burn front by fluid advection. In both cases however, the structure of the magnetic field close to the burn front is not significant, owing mainly to the fact that $\omega \tau_{bf}$ is small in all cases due to higher collisionality in this region. Additionally, the use of a uniform initial magnetic field profile reduces the importance of field transport towards the cold fuel, as will be discussed further in Section~\ref{sec:capburn}. A more detailed understanding of magnetic field transport effects in the burn front using this model is also the subject of future work.

\subsection{Regimes of Magnetized Burn}
To conclude this section, based on the results presented here, we can define three main regimes of magnetized burn:
\begin{itemize}
    \item \textbf{Thermal Conduction Driven Burn}: at low initial magnetizations the burn profiles are dominated by smoothing from thermal conduction, meaning larger burn scale lengths, and higher temperature fuel close to the burn front. A significant amount of alpha heating occurs and non-local deposition near the burn front is very important. Thermal conduction aids burn by preheating the cold fuel and allowing non-local alpha heating to deposit energy further into the cold fuel.
    \item \textbf{Alpha Heating Driven Burn}: at intermediate magnetizations thermal conduction is strongly suppressed, but alpha transport is not affected. Here, burn propagates only near the foot of the temperature profile, driven by non-local alpha heating, whilst thermal conduction can aid in this region due to lower $\omega \tau$. The hot fuel ($> 5$~keV) is effectively isolated from the burn front due to suppressed thermal conduction across the hotspot.
    \item \textbf{Strongly Suppressed Burn}: neither electron thermal conduction nor non-local alpha heating can transport heat out of the hot fuel region. Some burn can be driven through local alpha heating and radiation transport, however, burn progresses significantly more slowly.
\end{itemize}

\section{Magnetized High-Yield Capsule Simulations}
\label{sec:capburn}

\subsection{Magnetized Capsule Modeling using Chimera}

In this section, the effect of suppression of burn by a magnetic field will be studied in a 2D representative high-yield capsule simulation. Of particular interest is the coupled propagation of burn along magnetic field lines (along the pole of the capsule) and perpendicular to magnetic field lines (along the equator of the capsule), along with more realistic stagnation field profiles and hotspot temperature and densities.

The simulations presented were carried out in the Chimera radiation-magnetohydrodynamics code described in Section~\ref{sec:intro}, nominally based on the capsule and drive parameters used for NIF shot N210808 \cite{kritcherDesignInertialFusion2022} with a 40~T applied axial field. Capsule-only simulations were carried out using a frequency dependent radiation spectrum (FDS) applied as a boundary condition produced by hohlraum simulations carried out in the HYDRA radiation-hydrodynamic code \cite{marinakThreedimensionalHYDRASimulations2001}, with 67 frequency groups chosen to be sufficient to resolve both the thermal and m-band components of the radiation drive. These simulations were drive tuned in 1D to match experimental bang time, burn-averaged ion temperature, neutron yield and shock merger timing. This is achieved by use of a bulk drive energy multiplier of 0.96, along with a multiplier on the mass fraction of the surrogate dopant (germanium) of 4.213, compared to the tungsten dopant mass used in experiment. The process of drive tuning is carried out to ensure the regime of burn is comparable to current experiments. The use of drive multipliers in this way limits the predictive capability of these simulations, however in this work these simulations are only intended to be illustrative of the dynamics of magnetised burn, and any conclusions are drawn only by comparison to similarly tuned simulations. 

The simulations with an applied field were carried out in 2D in order to capture the expected anisotropy of burn propagation parallel and perpendicular to magnetic field lines, with the initial drive phase carried out on a spherical $r-\theta$ grid before being re-gridded onto a cylindrical $\rho-z$ mesh for stagnation and burn. It should be noted that no attempt was made to include realistic 2D perturbations, such as surface roughness or a fill tube defect, which means that the impact of high-mode instability growth and mix is not included in this modeling. Again, this limits the predictive capability of the simulations in respect to possible future magnetised ICF experiments, however this approach is chosen in order to better study the dynamics of magnetised burn propagation in a relevant geometry. In this case, a 2D simulation with no applied field achieves a neutron yield of 1.1~MJ and a burn-averaged ion temperature of 8.1~keV, which are similar to experimentally achievable regimes \cite{kritcherDesignInertialFusion2022}. Full details of implosion metrics for both baseline 2D burn on and burn off (i.e. no $\alpha$ deposition) simulations are given in Table~\ref{tab:cap_metrics0T}

\begin{table}
    \centering
    \caption{Implosion metrics for the baseline unmagnetized capsule simulations performed in Chimera. $t_{BT}$ is the bangtime (time of peak neutron production), $Y_N$ is the total DT neutron yield, $T_i$ is the burn averaged ion temperature, $\rho R_{tot}$ is the maximum shell areal density, $\Delta t_{burn}$ is the FWHM of the neutron production rate and $Y/Y_{no-\alpha}$ is the yield amplification with burn on.}
    \label{tab:cap_metrics0T}
    \begin{ruledtabular}
    \begin{tabular}{ccccccc}
         &$t_{BT}$&$Y_N$& $T_i$& $\rho R_{tot}$ &$\Delta t_{burn}$ &$\frac{Y}{Y_{no-\alpha}}$\\
 & (ns)& ($\times 10^{17}$)& (keV)& (g cm$^{-2}$)& (ns)&-\\
        \hline
        Burn off&9.12& 0.17& 4.2& 1.17& 0.19&-\\
        Burn on&9.21& 5.1& 8.1& 0.972& 0.12&30\\
    \end{tabular}
    \end{ruledtabular}
\end{table}

Simulations with the 40~T field include magnetic field transport with extended-MHD and anisotropic thermal conduction inline throughout the entirety of the implosion, however, the focus of the results presented in this paper will be purely on the stagnation phase. It is, however, important to appreciate the impact of magnetic field in the drive phase. Previous simulation \cite{walshPerturbationModificationsPremagnetisation2019,walshMagnetizedICFImplosions2022} and experimental \cite{moodyIncreasedIonTemperature2022,sioPerformanceScalingApplied2023} work on room-temperature magnetized platforms at the National Ignition Facility suggest that application of an axial magnetic field to a spherical capsule implosion can cause a change to the hotspot shape as measured by x-ray imaging. In particular, capsule-only simulations carried out in \citet{walshMagnetizedICFImplosions2022} show elongation of the hotspot along the direction of the applied field. Although beyond the scope of the current work, the primary cause of this perturbation is due to anisotropic heat flow, where magnetic suppression of heat flow at the capsule equator reduces the electron preheat at shock fronts, and leads to an increase to fuel compressibility. 

\begin{figure}
    \centering
    \includegraphics[width=3.37in]{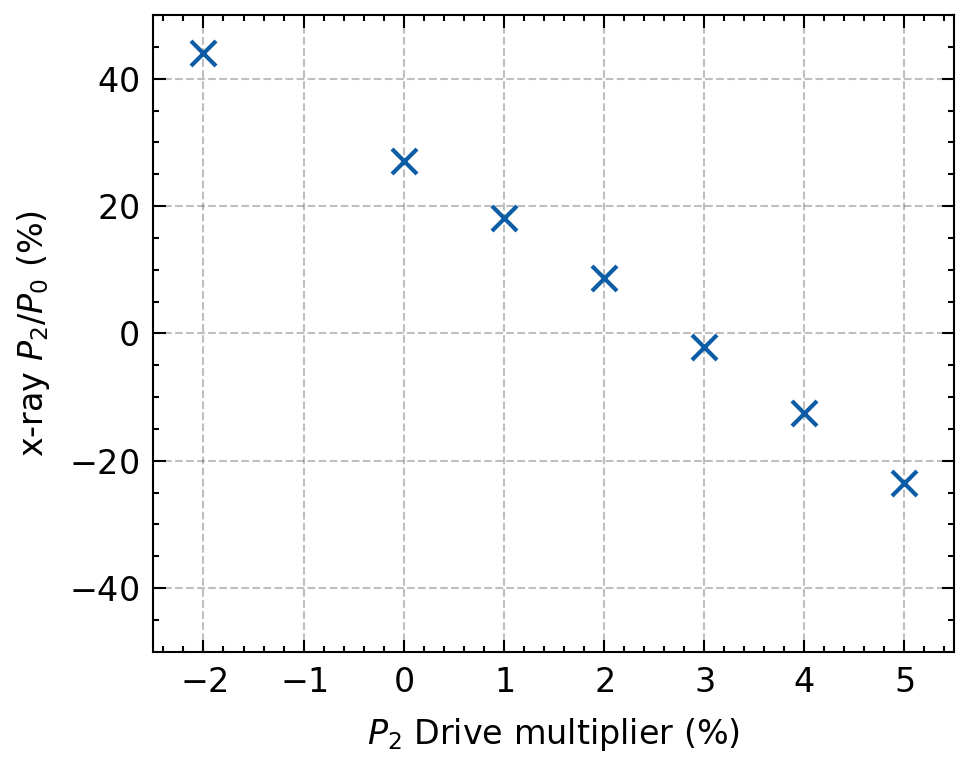}
    \caption{Measured $P_2/P_0$ from synthetic, time-integrated, fuel-only x-ray images as a function of applied $P_2$ radiation drive multiplier for a set of burn off simulations with a 40~T applied field. }
    \label{fig:cap_p2tune}
\end{figure}

In the simulations performed in this work, the addition of the 40~T axial field demonstrates this elongation of the hotspot along field lines, with a measured x-ray $P_2/P_0 = 27\%$ at neutron bangtime for the burn off case.  In this study, it is desirable to mitigate the impact of this drive phase perturbation, and this is achieved here through the addition of a uniform mode-2 asymmetry to the x-ray drive to compensate for the increased compression along the capsule equator. Figure~\ref{fig:cap_p2tune} shows the variation of x-ray $P_2$ as a function of radiation drive $P_2$ for a set of magnetized burn-off simulations, indicating that a multiplier on the radiation source $P_2 \sim 3\%$ produces a hotspot that is close to round. Note that this $P_2$ multiplier is applied to the FDS source at the simulation boundary, and diffusion of radiation due to the $P_{1/3}$ radiation transport scheme employed leads to a lower drive asymmetry at the ablation surface. In addition, it should be noted that this type of tuning is not representative of hohlraum dynamics, and the resultant performance of the tuned capsule simulations may not be indicative of the best performing magnetized capsule set-up, where, for example, a time-dependent $P_2$ source may perform better and more accurately capture hohlraum physics. For the remainder of this paper, only the cases tuned to mitigate drive phase mode-2 will be considered.

\begin{table}
    \centering
    \caption{Implosion metrics for the tuned 40~T magnetized capsule simulations performed in Chimera. $t_{BT}$ is time of peak neutron production, $Y_N$ is the total DT neutron yield, $T_i$ is the burn averaged ion temperature and $P_2/P_0$ is measured from time integrated x-ray images.  $Y/Y^{B = 0}$ and $T_i/T_i^{B=0}$ are the fractional yield and temperature changes from the corresponding unmagnetized simulation.}
    \label{tab:cap_metrics40T}
    \begin{ruledtabular}
    \begin{tabular}{ccccccc}
         &$t_{BT}$&$Y_N$& $T_i$& $P_2/P_0$&$\frac{T_i}{T_i^{B=0}}$&$\frac{Y}{Y^{B = 0}}$\\
 & (ns)& ($\times 10^{17}$)& (keV)& & &\\
        \hline
        Burn off&9.12& 0.22& 5.3& -0.02& 1.26&1.31\\
        Burn on&9.20& 2.9& 8.1& +0.14& 1.00&0.58\\
    \end{tabular}
    \end{ruledtabular}
\end{table}

A summary of select implosion metrics are given in Table~\ref{tab:cap_metrics40T} for the 40~T magnetized capsule simulations with both burn on and burn off. In the case with burn off, both ion temperature and fusion yield are enhanced by $\sim 30\%$ with a 40~T applied field. Comparison can be made with scaling model for these metrics in the presence with an applied field derived in \citet{walshMagnetizedICFImplosions2022}, which predicts temperature amplification of 1.37 and yield amplification of 1.6 in the limit of high $\omega \tau$ in the stagnated hotspot. This agrees moderately well with simulations, with the discrepancy likely caused by the decrease in simulated performance due to shape effects, in addition to effects not captured in the scaling model such as the topology of the stagnated B-field, as will be discussed in more detail in Section~\ref{subsec:hotspot}. 

The case with burn on shows a substantial decrease in performance ($\sim 40\%$), indicating that there are fundamental changes to the onset of ignition and burn propagation in this magnetized capsule design, that will be studied in detail in the remainder of this paper. Ion temperature is the same in both 0~T and 40~T cases, however, the causes for the measured ion temperature differ in each case. In the 0~T case, the ion temperature increases substantially when burn is on through bootstrap heating, which is correlated to overall fusion yield. In the magnetized case, the bootstrap heating is lower due to decreased fusion performance, however, the ion temperature is raised in comparison to a similarly performing 0~T case due to magnetic insulation of the hotspot.

It should be noted here that the performance decrease in the presence of a magnetic field is only representative of the single simulation set-up carried out in this work, and may not apply broadly to high-yield magnetized target designs. In particular, these simulations are simplistic in comparison to experiments, neglecting realistic drive effects (time-dependent $P_2$, $P_4$ etc.) and the impact of high-mode instability growth (which magnetic fields have been predicted to suppress \cite{walshPerturbationModificationsPremagnetisation2019}). Additionally, due to the nature of the so-called `ignition cliff' regime in which these simulations lie, small perturbations to the onset of ignition can lead to large changes in obtained yield. Detailed comparison of the MHD code Chimera to experiments in this regime are yet to be carried out, limiting the predictive capability of these simulations on metrics such as yield. In the remainder of this work these simulations are predominantly used to investigate the physical processes of ignition and burn, which can still be fully understood despite the limitations outlined here. 



\subsection{Hotspot formation with a magnetic field} \label{subsec:hotspot}

\begin{figure}
    \centering
    \includegraphics[width=3.37in]{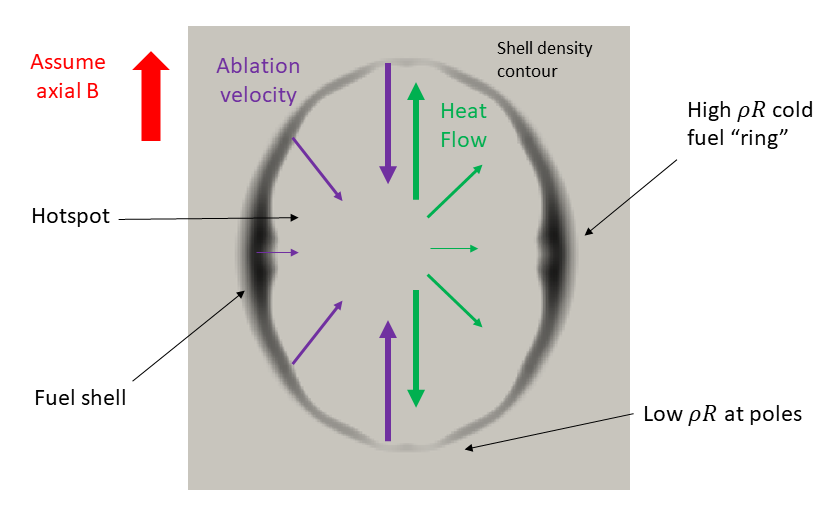}
    \caption{Schematic showing the impact of a strong axial magnetic field on ablation velocity during hotspot formation, overlayed on a representative fuel density contour from the 40T simulation (grayscale). The size of the green arrows represent the magnitude of electron heat flux, whilst the size of the purple arrows represents the magnitude of the ablation velocity from the cold fuel shell.}
    \label{fig:cap_ablation}
\end{figure}

During stagnation, ignition and burn, ablation of material from the cold fuel layer into the hotspot is significant, leading to an increase in hotspot mass and a corresponding reduction in shell density. This ablation is driven by energy transport into the cold fuel; at early times the energy transport is driven mainly by thermal conduction, whilst at later times, close to ignition, alpha heating begins to become an important driver of ablation. The presence of a magnetic field, sufficient to magnetize electron thermal conduction, will therefore be expected to change the rate of ablation of cold fuel into the hotspot. 

The diagram in Fig. \ref{fig:cap_ablation} illustrates the change to ablation velocity from the dense fuel shell as a function of polar angle caused by the applied field. Here, we assume initially that the applied field at stagnation is purely axial, with electron Hall parameter $\omega \tau > 1$ everywhere, sufficient to substantially modify the heat flux into the cold fuel. The magnitude of the heat flow towards the cold fuel reduces moving from pole to equator, leading to a reduction in mass ablation velocity. This leads to a higher areal density ring forming at the equator, as observed in the density contour from the simulation, with a thinning of the fuel shell along the polar direction.

\begin{figure*}
    \centering
    \includegraphics{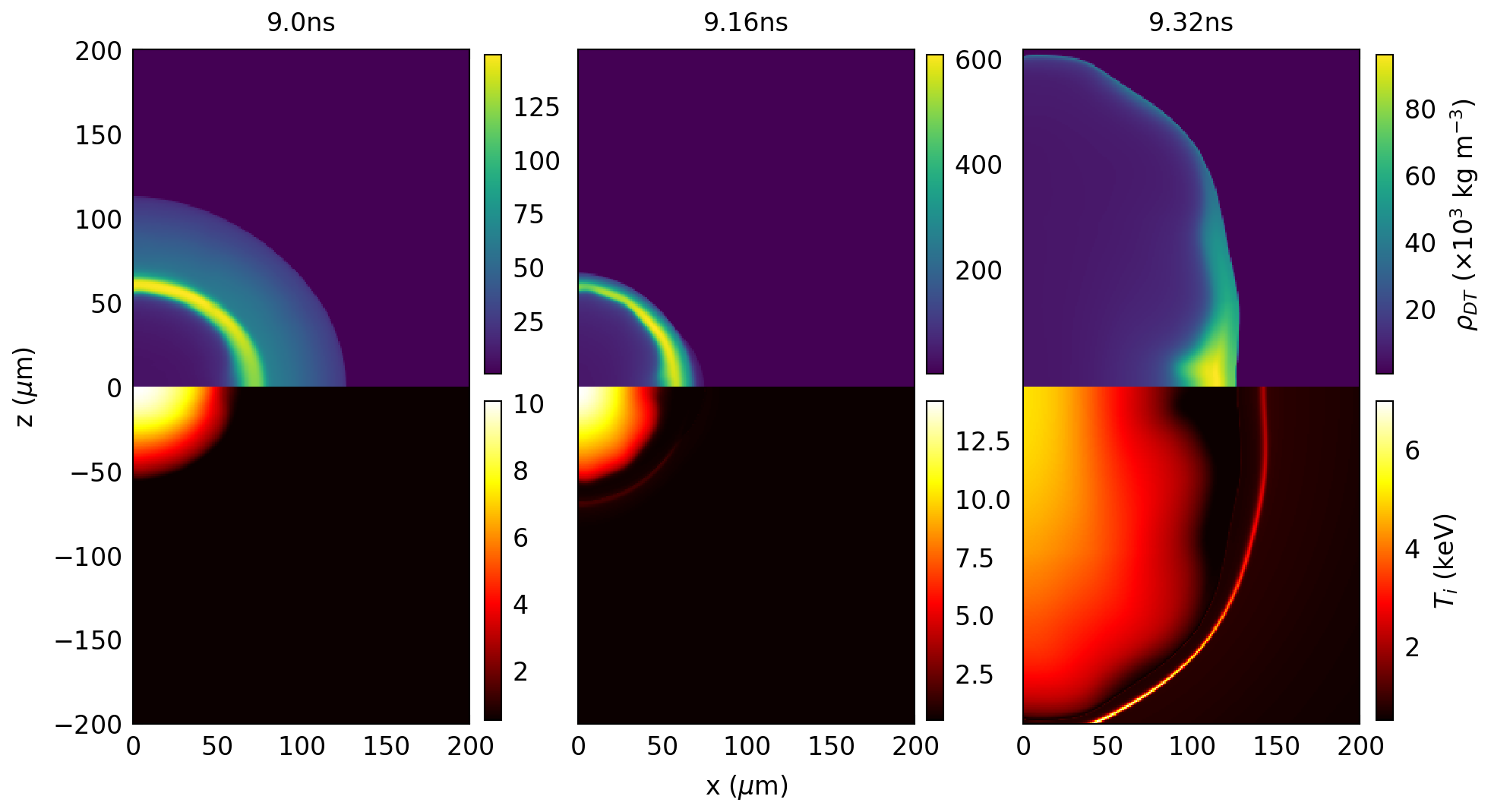}
    \caption{Fuel density (top half) and ion temperature (bottom half) profiles for the 40~T capsule simulation at three representative times during stagnation and burn: the start of stagnation (9.0~ns); the time of peak compression (9.16~ns); and the end of burn propagation (9.32~ns).}
    \label{fig:cap_2D-profiles}
\end{figure*}

Fig. \ref{fig:cap_2D-profiles} shows fuel density and ion temperature profiles for the 40~T magnetized capsule simulation at three separate times. The first panel shows the capsule shape during stagnation, where the free-falling shell is visible surrounding the denser inner shell out to $\sim 120$ $\mu$m. The inner shocked shell has a noticeable oblate shape due to the tuned radiation drive $P_2$.  The profiles at 9.16~ns correspond to the time of peak compression (the minimum of volume integrated $PdV$ power), noting that this occurs earlier than the time of peak neutron production ($\sim 9.2$~ns) due to the large amplification in neutron production during burn propagation. By this time the formation of a cold fuel ring at the equator is visible due to reduced ablation at the capsule equator. The overall shape of the fuel shell is close to round, compared to the oblate shape at 9.0~ns, indicating that this tuned simulation is close to maximising hydrodynamic implosion efficiency.

The profiles shown for 9.32~ns show a time at the end of burn propagation, where total neutron production has fallen to $< 1\%$ of the maximum rate. Here, the fuel assembly is substantially elongated along the applied field direction, showing a rapid change in shape from stagnation. This is predominantly driven by different burn propagation rates at the capsule equator and the pole. At the equator, there is a large remaining cold fuel mass which cannot be efficiently burned due to quenching of hotspot temperature by rapid burn along the pole. The rapid change of shape in this magnetized igniting capsule is symptomatic of burn in a magnetized spherical capsule.

\begin{figure}
    \centering
    \includegraphics[width=3.37in]{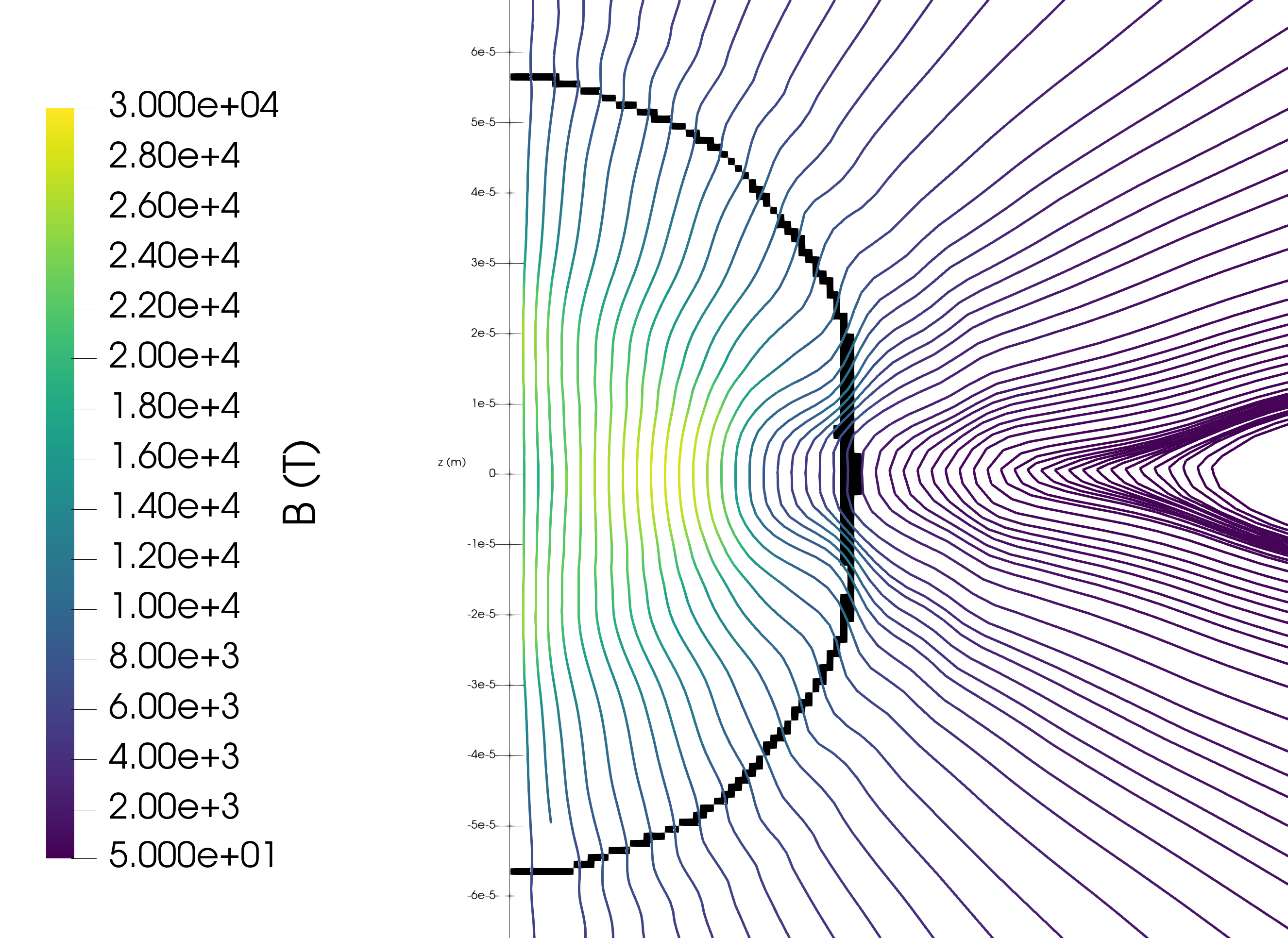}
    \caption{Magnetic field streamlines for the 40~T burn on simulation at 9.16~ns. The black contour shows the position of the 2~keV ion temperature contour. The direction of the field lines point vertically upwards.}
    \label{fig:cap_2DB}
\end{figure}

\begin{table}
    \centering
    \caption{Representative values for hotspot magnetization at stagnation for the 40~T implosion simulation. $B_{max}$ is the peak magnetic field magnitude, $\omega\tau_{max}$ is the peak Hall parameter, $\overline{\omega\tau}$ is the burn-averaged Hall parameter, $\omega\tau_{2keV}$ is the Hall parameter at the 2~keV ion temperature contour as measured at the capsule equator, and $\beta_{min}$ is the minimum value of the plasma beta (ratio of thermal to magnetic pressure).}
    \label{tab:cap_field}
    \begin{ruledtabular}
    \begin{tabular}{cccccc}
         &$B_{max}$ (kT)&$\omega \tau_{max}$& $\overline{\omega \tau}$ &$\omega \tau_{2keV}$& $\beta_{min}$\\
        \hline
        Burn off&40.2& 23& 13 &0.54& 40\\
        Burn on&28.3& 52& 22 &0.24& 49\\
    \end{tabular}
    \end{ruledtabular}
\end{table}
The magnetic field profile close to stagnation (9.16~ns) is shown for the burn on simulation in Figure~\ref{fig:cap_2DB}, with representative quantities for fuel magnetization for both burn on and burn off cases in Table~\ref{tab:cap_field}. The overall shape of the magnetic field is in line with expectations for ideal flux compression for an initially axial field, with maximal compression at the capsule equator ($\perp$ to field) and no compression along field lines, producing a characteristic `hour glass' magnetic field profile. Magnetic field magnitudes are strongly peaked in the center of the hotspot where flux compression is the highest, with peak fields of $\sim 30 - 40$~kT possible during stagnation, corresponding to peak $\omega \tau \sim 50$. The Hall parameter remains large across the hotspot with $\omega \tau \sim 10$ a reasonable value to assume, corresponding to almost complete suppression of perpendicular heat flow. The magnetization drops rapidly at the hotspot edge, with $\omega \tau \sim 0.2$ at the 2~keV contour, due to the lower temperature and density here. These parameters mean that we expect burn propagation to be in the alpha driven regime as identified in Section~\ref{sec:planar_burn}.

Here, all magnetic transport effects during the implosion leading to the specific magnetic field profile observed are not studied in detail, and values quoted here are representative as the magnitude of the magnetic field changes rapidly due to the large change in fluid velocity close to stagnation. The burn on case has lower peak field magnitude due mainly to the increased hotspot thermal pressure caused by alpha heating, which creates a less compressed hotspot. In general, despite the large magnetic field magnitudes, the enormous thermal pressure in ICF hotspots dominates the magnetic pressure, with minimum plasma $\beta \sim 50$, and typically $\beta > 1000$ away from the peak in magnetic field near the hotspot center, due to the $B^2$ scaling of magnetic pressure.

This more complicated hour glass magnetic field profile demonstrates an oversimplification in the previous discussion on magnetized capsule implosions, where we consider the two extreme cases of capsule pole and capsule equator. In actuality, magnetic field lines are generally more perpendicular to the hotspot surface across a broader range of polar angles than would be for a purely axial field. Additionally, magnetic field lines here appear to wrap around the cold fuel ring forming at the capsule equator, which will restrict thermal transport into this feature from all angles, effectively insulating it further. It is important, therefore, to appreciate this 2D picture of the magnetized hotspot for the discussion that follows. 

\subsection{Burn propagation in a magnetized capsule} \label{subsec:burn}

\begin{figure}
    \centering
    \includegraphics[width=3.37in]{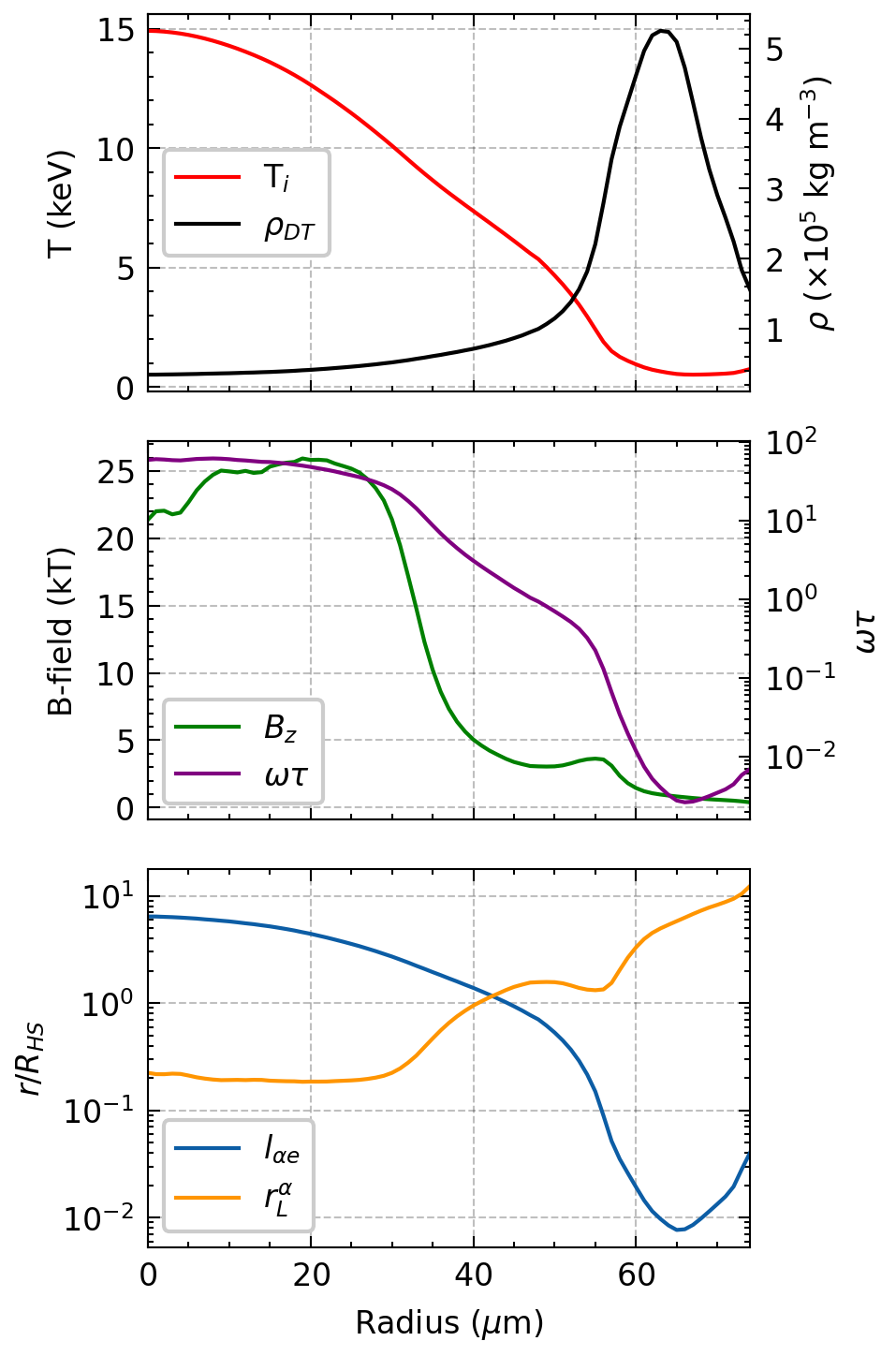}
    \caption{Lineouts along the capsule equator for the 40~T capsule simulation at 9.2~ns showing: (Top) Ion temperature and fuel density; (Center) magnetic field and electron Hall parameter; (Bottom) Birth $\alpha$ stopping length and Larmor radius normalized to hotspot radius.}
    \label{fig:cap_mag-profiles}
\end{figure}

The effect of an axial magnetic field on burn propagation in a magnetized ICF capsule can be better understood by comparing the 1D dynamics of burn along the capsule equator (fully $\perp$ to the field) and along the pole (fully $\parallel$ to the field), and comparing to the results presented in Section \ref{sec:planar_burn} in planar geometry.

1D lineouts along the capsule equator from the 2D 40~T axial field simulation are shown in Figure \ref{fig:cap_mag-profiles} at 9.20~ns, which corresponds to the time of peak neutron production (i.e. peak burn). The top panel shows ion temperature and fuel density, with a notable step in the ion temperature profile at $\sim 30$~$\mu$m, indicating that burn perpendicular to field lines in the full capsule simulation is in the alpha driven regime identified in Section \ref{sec:planar_burn}. 

The center panel of Figure~\ref{fig:cap_mag-profiles} shows the perpendicular component of the magnetic field ($B_z$) alongside the electron Hall parameter. The magnetic field profile is strongly peaked towards the center of the capsule, with values of $B \sim$ 25~kT at radius $\lesssim$ 30~$\mu$m. The field profile in the central region is complex, with it being predominantly set during the time period in which the shock reflects off the axis, coupled with bulk hotspot dynamics during stagnation. At $r > 30$~$\mu$m the field magnitude falls rapidly, with a magnitude of $B \sim 3 - 4$ ~kT across the remainder of the hotspot. This is consistent with advection of the field into the central hotspot as cold fuel is ablated into the hotspot. Close to the hotspot edge there is a secondary peak in magnetic field magnitude which is observed to be produced due to the Nernst effect, which causes an advection of magnetic field down the steep temperature gradients close to the hotspot edge.  The Hall parameter profile shows a similar rapid variation across the hotspot region. In the central 30~$\mu$m, Hall parameters are high ($> 10$) resulting in a suppression of thermal conductivity to $< 1\%$ of the unmagnetized value. Across the remainder of the hotspot the Hall parameter remains $> 1$, which indicates that there will still be significant suppression of thermal transport due to the magnetic field. Moving into the cold fuel region, temperature falls and density increases, with a Hall parameter at the 2~keV contour $\sim 0.1$, indicating that electron thermal conduction will not be significantly restricted at the burn front.

The lower panel of Figure \ref{fig:cap_mag-profiles} plots the Larmor radius and collisional mean free path for a birth alpha particle as a function of radius, with both normalized to the approximate hotspot radius ($R_{HS} = 50$~$\mu$m). In the strongest magnetic field region ($r < 30$~$\mu$m), the alpha particle Larmor radius is approximately 2 orders of magnitude smaller than the alpha stopping distance, indicating that alpha particle transport is restricted by the presence of the field. Additionally, $r_L^\alpha \sim 0.5R_{HS}$, meaning that alpha particles born close to the hotspot center and moving purely in along the $B_\perp$ direction are confined within the central strong field region. It should be noted that particles with any velocity component towards the capsule pole can escape this strong field region. As the magnetic field strength falls rapidly beyond 30~$\mu$m, $r_L^\alpha$ increases significantly, whilst increased density and decreasing temperatures mean that $l_{\alpha e}< r_L^\alpha$ in this region. This indicates that local alpha heating within the hotspot center will be enhanced by the presence of the strong central field, whilst alpha particles escaping the innermost strong field region will slow and deposit their energy within the colder fuel near the hotspot edge, as in typical unmagnetized simulations.

\begin{figure*}
    \centering
    \includegraphics{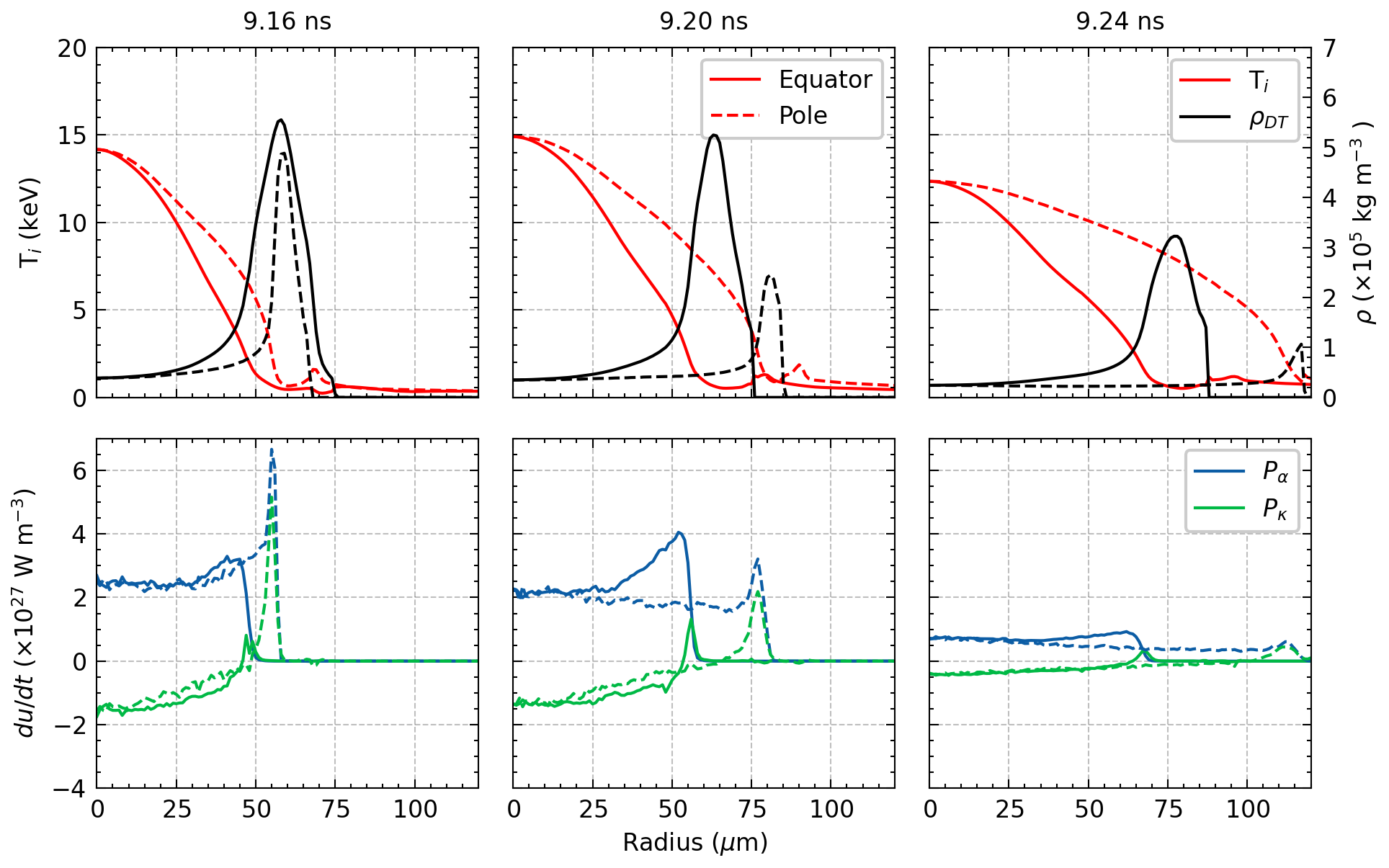}
    \caption{Profiles along capsule equator (solid) and pole (dashed) for the 40~T axial magnetized capsule simulation for three times covering burn propagation. Top panels show ion temperature in red and fuel density in black. Bottom panels show power density from alpha heating in blue and electron thermal conduction in green.}
    \label{fig:cap_burn-profiles}
\end{figure*}

Figure \ref{fig:cap_burn-profiles} shows the evolution of ion temperature and DT density profiles along both the capsule equator and capsule pole from peak compression (9.16~ns), during peak burn (9.2~ns), and at the beginning of capsule disassembly (9.24~ns). This time period accounts for 67\% of the total neutron production during the implosion. These profiles are plotted alongside the power density contributions from alpha heating and electron thermal conduction, allowing the results to be contrasted to results in Section \ref{sec:planar_burn}.

It is important to note that burn propagation along equator and pole can not be decoupled entirely from the anisotropy caused by the field during hotspot formation. Looking at the ion temperature and fuel temperature profiles at 9.16~ns, the total mass in the fuel shell is is smaller, with both a narrower and lower magnitude peak in DT density, due to enhanced mass ablation rate during stagnation. This also leads to a broader temperature profile, with the 2~keV contour extending to 56~$\mu$m in radius at the capsule pole, compared to 48~$\mu$m at the capsule equator. The power balance profiles indicate that burn is already propagating in the polar direction, with a significant peak in alpha heating at the hotspot edge $\sim 2$ times larger than the magnitude of alpha power deposition at the center of the hotspot. Along the capsule equator, the onset of rapid burn propagation due to non-local alpha heating has not occurred by 9.16~ns, evidenced by the lack of a significant peak of alpha deposition at the hotspot edge. This is due in part to the denser, colder fuel region close to the hotspot edge providing a broader region in which alpha energy is dissipated. Additionally, the high magnetic field strengths in the inner hotspot will restrict alpha transport in this direction. The onset of burn propagation along the polar direction prior to peak compression may have a negative impact on capsule performance, as the enhanced pressure wave ahead of the burn front will lead to reduced compression along the polar direction. 

The profiles at 9.2~ns demonstrate the two contrasting regimes of burn propagation in a magnetized capsule. Along the polar direction burn propagation is much more rapid, with the 2~keV contour now extending beyond 75~$\mu$m. This measurement is due to both enhanced propagation of the front and by the rapid re-expansion of the fuel shell driven by burn, with the peak in fuel density at a significantly larger radius along the polar direction. At the capsule equator, burn propagation has onset, with a significant non-local alpha heating peak observed in the power balance profiles. The ion temperature profile has a notable stepped shape, with an inflection at $\sim 30$~$\mu$m, which indicates that burn along the capsule equator is qualitatively similar to the alpha driven burn regime identified in Section~\ref{sec:planar_burn}. The step in the ion temperature profile corresponds to the region of rapid drop-off in magnetic field strength observed in Figure~\ref{fig:cap_mag-profiles}. Alpha energy deposition is uniform within the central hotspot, with a broader non-local alpha heating peak observed beyond 30~$\mu$m, indicating that alphas escaping the inner region are relatively non-local and can propagate burn into the cold fuel. These results indicate that the dominant cause of reduced burn propagation rate perpendicular to magnetic field lines is the restriction of energy transport within the hottest part of the fuel.

The final profiles in Figure~\ref{fig:cap_burn-profiles} show the conditions towards the end of neutron production. Here, central hotspot temperature is dropping from its peak value at 9.2~ns, with the rapid re-expansion along the pole of the capsule leading to a quenching of burn. It is clear from the ion temperature and fuel density profiles that, by a combination of rapid burn propagation and capsule disassembly, the dense fuel along the capsule pole has been mainly ablated into the hotspot. Profiles along the capsule equator show that burn propagation is still progressing in the alpha driven regime at a lower rate. However, due to the falling central temperatures, this burn wave can not be sustained long enough for a significant amount of burn in the equatorial direction. 

\begin{figure}
    \centering
    \includegraphics[width=3.37in]{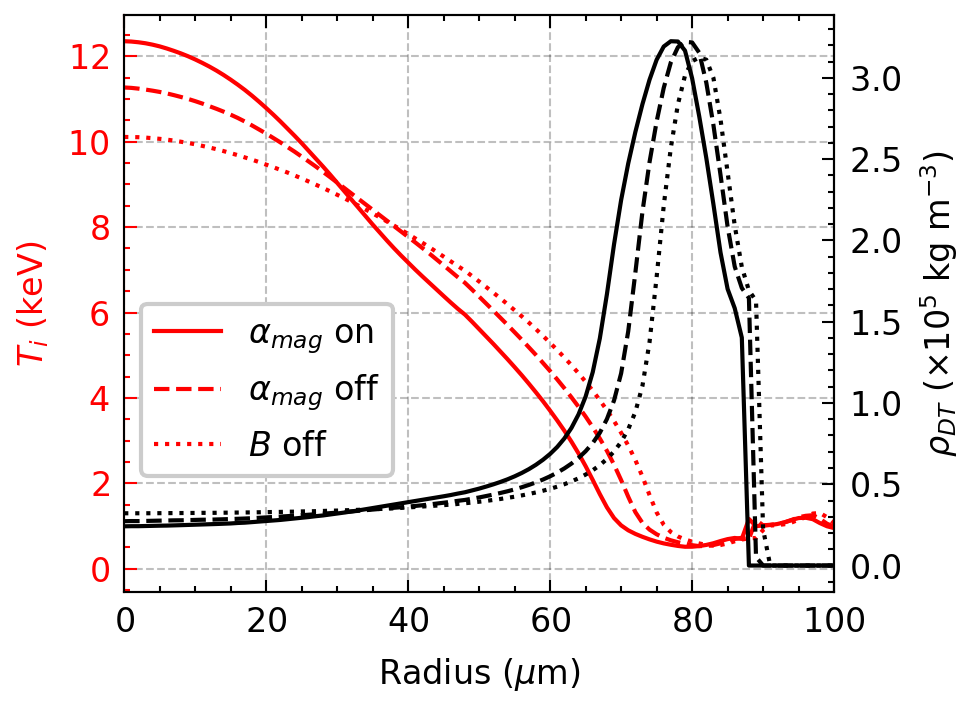}
    \caption{Equatorial line-outs showing ion temperature and fuel density at 9.32~ns for 3 simulations of the high-yield capsule with a 40~T axial field: a case with full MHD (solid); $\alpha$ particle magnetization effects switched off at 9.14~ns (dashed); and all MHD effects switched off at 9.14~ns (dotted).}
    \label{fig:cap_B-effects}
\end{figure}

The separate impacts of magnetized electron thermal conduction and magnetized $\alpha$ transport can be tested by re-initializing the magnetized simulation close to the time of peak compression (9.14~ns) with aspects of the MHD package in Chimera switched off. In this paper, we study two such cases: one with the field present but unmagnetized $\alpha$ transport; and one with the field removed entirely at the onset of burn.  The equatorial ion temperature and fuel density profiles at 9.24~ns for each of these simulations are compared in Fig.~\ref{fig:cap_B-effects} in order to study the maximal impact of each mechanism on burn propagation.

Comparison of the ion temperature profiles for the 3 cases considered show that burn propagation rate increases with removal of MHD effects, with the position of the 2~keV contour measured to be at 66~$\mu$m, 70~$\mu$m and 73~$\mu$m for the fully magnetized, unmagnetized $\alpha$ and no field cases respectively. By comparison with the results for the planar model presented in Sec.~\ref{sec:planar_burn}, the stepped temperature profile associated with alpha driven burn becomes less significant with removal of MHD effects. With the removal of magnetized alpha particle transport the inflection remains, but the gradient of the inflection region is less steep. In the case with the magnetic field removed, the ion temperature gradient is now smooth across the hotspot region, indicating a return to the conduction driven burn regime identified in Sec.~\ref{sec:planar_burn}.

\begin{figure}
    \centering
    \includegraphics[width=3.37in]{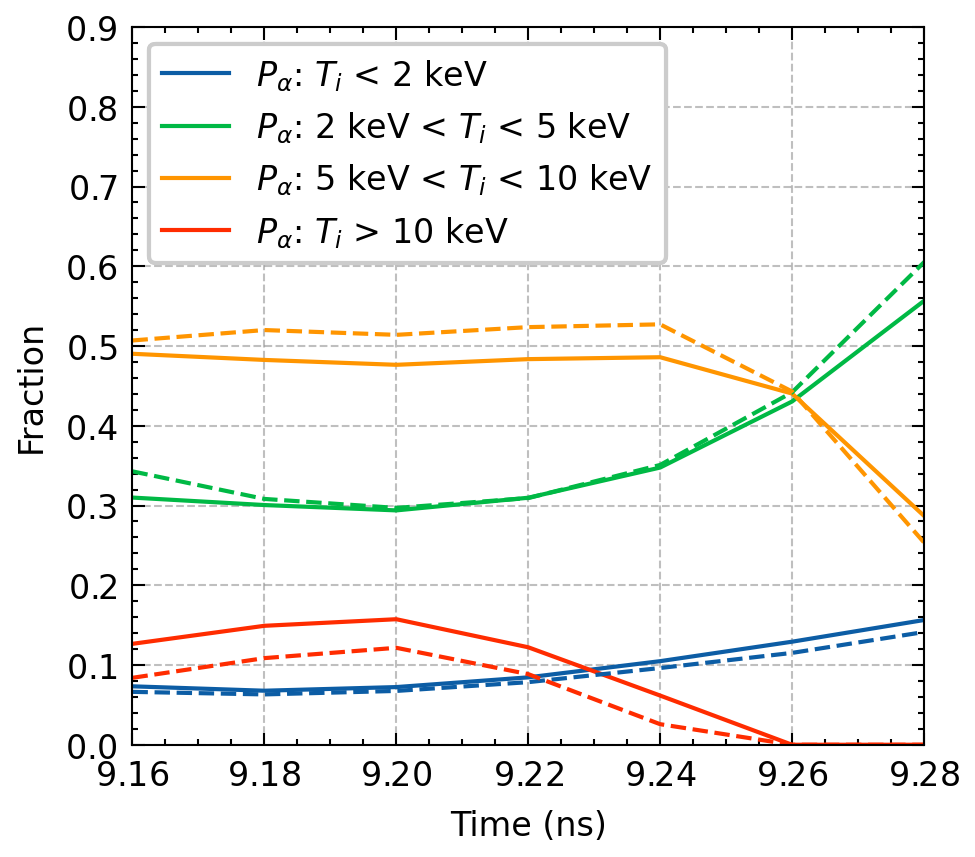}
    \caption{Plots of the total volume-integrated alpha power deposition in fuel within the given ion temperature ranges for the simulation with full MHD effects (solid) and with $\alpha$ magnetization switched off at 9.14~ns (dashed), expressed as a fraction of total $\alpha$ power deposition as a function of time over the main burn pulse. }
    \label{fig:cap_alpha-dep}
\end{figure}
The role of the magnetic field on $\alpha$ deposition is illustrated further in Fig.~\ref{fig:cap_alpha-dep}, which shows the fraction of total volume integrated alpha power deposition within fuel of a given ion temperature range for the simulations with and without magnetized $\alpha$ trajectories across the main burn phase. The only notable effect of magnetizing $\alpha$ transport is to modify the deposition of alpha energy in the hottest fuel ($T_i > 5$~keV). Here, the fraction of alpha energy deposited in fuel with $T_i > 10$~keV is consistently $\sim 5\%$ higher with B-field effects present across the main burn pulse. This leads to a corresponding reduction in the proportion of alpha energy deposited in fuel $5 < T_i < 10$~keV. This can be expected to have a reasonably significant effect on burn in a magnetized capsule as this is the fuel region in which $\sim 50\%$ of the total alpha energy is deposited.

These results highlight a key difference between burn in the integrated capsule simulations compared to the planar model, in that the magnetic field strength is not uniform across the hotspot, but rather peaked at radii $\lesssim 30$~$\mu$m. This forms an efficient barrier to both thermal and $\alpha$ transport leading to a reduced rate of energy transport towards the hotspot edge. This acts to reduce the rate of burn propagation primarily by lowering the fuel temperature at larger radius, hence restricting the energy available to drive burn propagation, in addition to reducing the rate of alpha production close to the cold fuel interface.

\section{Conclusions}
\label{sec:conc}
The propagation of a thermonuclear burn wave is studied in a magnetized ICF capsule, using an idealized model in planar geometry in Section~\ref{sec:planar_burn}, before burn in a magnetized high-yield capsule is studied in Section~\ref{sec:capburn}. The planar model allows a qualitative understanding of the dynamics of burn propagation in the direction perpendicular to magnetic field lines. This leads to the definition of three distinct regimes of burn propagation: conduction driven burn, where thermal conduction plays an important role in transporting energy to the burn front; alpha driven burn, where thermal conduction to the burn front is suppressed, but non-local alpha heating can still drive burn propagation; and fully suppressed burn, where energy transport to the burn front by both mechanisms is suppressed by the presence of the field. 

For the capsule simulations in Section~\ref{sec:capburn} it is noted that the magnetic field impacts spherical ICF implosions by various mechanisms across the entirety of the simulation. In this paper, efforts have been made to restrict the study only to the effect of the magnetic field on the stagnation and burn phases of the capsule simulation. This means that many of the specific impacts of the magnetic field are not studied in detail here, in particular its effect on bulk implosion shape, which primarily appears to occur due to modifications to incoming shock propagation speed perpendicular to field lines. A major limiting assumption in this work is that this effect can be accounted for by using a simple uniform multiplier on mode-2 radiation drive shape. This limits the predictive capability of these simulations in terms of fully integrated metrics, such as nuclear yield, which is observed to decrease in the simulations carried out. However, the simulations here do give a qualitative indication as to the physical effects of a magnetic field on burn propagation. Simulations of an igniting NIF capsule design with a 40~T applied magnetic field reach a peak field strength of 27~kT at stagnation, with Hall parameters > 50 observed. These field strengths are also capable of restricting alpha particle range in the hotspot, with $\alpha$ Larmor radii < $0.5 R_{HS}$ observed. Prior to the onset of burn propagation, the shape of the imploded fuel shell is modified by the presence of the field, as reduced thermal conductivity at the equator lowers the ablation rate of the cold fuel layer during stagnation, leading to a ring of colder, denser fuel to form at the capsule equator. Burn propagation in the magnetized capsule is highly anisotropic, with the bulk implosion shape varying rapidly over the $\sim 300$~ps main burn pulse, from a slight oblate shape prior to stagnation to a large measured prolate shape ($P_2/P_0 \sim 50\%$) at 9.32~ns. The cause of this is studied in detail, by comparison of burn along the capsule equator ($\perp$ to field lines) and pole ($\parallel$ to field lines), where it is seen that burn is in the alpha driven regime at the capsule equator. Correspondingly, higher central temperatures due to reduced heat losses at the equator can drive more rapid burn propagation along the polar direction. 

This point highlights a key conclusion in this paper, in that the dynamics of an axially magnetized capsule must be modeled as a 2D system at minimum to accurately capture the impact of the field. Ultimately, any suppression of burn in the equatorial direction can be recycled as additional burn propagation along field lines, meaning that suppressed burn need not be entirely detrimental to yield. In the case of the simulation presented here, the total yield is limited by the rapid burn achieved along the capsule pole, which quenches burn propagation before a substantial amount of the fuel at the capsule equator can be burnt. This set-up is likely far from an ideal implosion design to be magnetized, as the capsule already ignites without a field present, meaning the primary benefit of the field (lower ignition threshold) isn't best utilized. A better design may include higher $\rho R$ shell to provide additional confinement along the pole, and allow further time for burn propagation along the equator, where the addition of the field also allows for ignition when it otherwise may not occur. Further study should be undertaken to investigate in detail how shape tuning may affect overall yield, such as including time-dependent $P_2$ (for example in fully-integrated hohlraum and capsule simulations).  This work does not study the potential benefit of the stabilization of high-mode instability growth perpendicular to magnetic field lines \cite{walshMagnetizedAblativeRayleighTaylor2022}, instead opting to consider only low-mode perturbations in order for comparison with 1D models. As mix is one of the primary degradation mechanisms in current NIF high-yield implosions \cite{reganHotspotMixIgnitionscale2012}, any potential beneficial impact of the field may be highly important to increasing the obtained fusion yield, and should be a focus for future integrated magnetized capsule simulations.

In conclusion, the suppression of burn propagation perpendicular to magnetic field lines in magnetized ICF capsules relevant to current NIF designs has been demonstrated in MHD simulations of a high-yield capsule with a 40~T axial field. It is clear from these results that optimizing fusion performance in a magnetized capsule will require many changes from current high-performing NIF designs. To this end it is important to further explore capsule design space, particularly regions not previously considered, in which unmagnetized capsules may not be able to ignite and the beneficial effect of the field can be maximized. Magnetized designs could also be further optimized by considering alternative initial magnetic field topologies. One idea is to obtain closed field lines in the stagnated hotspot, which will maximize the suppression of thermal losses from the hotspot. Whilst the results here may suggest this would be detrimental for burn rates, more isotropic burn could allow for longer burn timescales by avoiding the quenching of hotspot temperature by rapid burn along a preferential direction. Ultimately, performance of a magnetized ICF capsule will depend on many factors, impacting choices of laser drive, hohlraum design, field topology and capsule design, motivating further study both numerically and experimentally.


%
%

%


\section*{Acknowledgments}
Work was partly performed under the auspices of the U.S. Department of Energy by Lawrence Livermore National Laboratory under Contract No. DE-AC52-07NA27344, LDRD projects 20-SI-002 and 23-ERD-025. 

\section*{Data Availability}
The data that support the findings of this study are available from the corresponding author
upon reasonable request.

\bibliography{references.bib}

\end{document}